\documentclass[nofootinbib,aps,11pt]{revtex4-2}
\usepackage{mathtools}
\usepackage{amssymb}
\usepackage{amsthm}
\usepackage{graphicx}
\usepackage{subcaption}
\usepackage{caption}
\usepackage{makecell}
\usepackage{hyperref}
\usepackage{cancel}
\usepackage{amssymb}
\usepackage{textcomp}
\usepackage{amsmath}
\usepackage{bm}
\usepackage{times}
\usepackage{epsfig}
\usepackage{color}
\usepackage{float}
\usepackage{multirow}
\usepackage[normalem]{ulem}
\usepackage{comment}

\begin{document}
	\title{Gauge Interactions Involving New Vector and Pseudoscalar Particles}
	\bigskip
	\author{Ruiyu Xing}
	\author{Honglei Li$^1$}
	\email{sps\_lihl@ujn.edu.cn}
	\author{Zhi-Long Han$^1$}
	\email{sps\_hanzl@ujn.edu.cn}
	\author{Fei Huang$^1$}
	\email{sps\_huangf@ujn.edu.cn}
	\author{Cui-Juan Lv}
	\email{sps\_lvcj@ujn.edu.cn}
	\author{Xinyi Zhang$^1$}
	\email{sps\_zhangxy@ujn.edu.cn}
\affiliation{School of Physics and Technology, \\ 
University of Jinan, 250022, Jinan, China}
	\begin{abstract}
		The Vector gauge bosons $Z'$ and axion-like particles (ALPs) $a$ are predicted by the $U(1)_X$ gauge extension and Peccei–Quinn (PQ) models, respectively. We focus on the interactions of the $Z'$ boson and the $a$ particle with the $W^\pm$ gauge boson, and investigate the production cross section and related kinematic distributions for the $Z'/a\to W^+W^-$ decay process at hadron colliders. We consider the angular distributions of charged leptons in the final states corresponding to the semileptonic and fully leptonic decays of the $W^\pm$ boson, separately. The study shows that, in the associated production processes of $Z'/a$ with $W^+$ or $Z$ bosons, the angular distributions of final-state leptons can serve as an important observable for probing new gauge interactions, and can also provide a criterion to distinguish vector particles from pseudoscalar particles.
        
	\end{abstract}
	\maketitle
	\newpage
	\quad
	\section{Introduction}
	New physics beyond the Standard Model (SM) has always been a fascinating subject in particle physics research. Gauge symmetry is the cornerstone of quantum field theory, and numerous new theories have been proposed to extend the gauge interactions. These theories are of great help in addressing issues such as dark matter~\cite{1980ApJ,Planck:2018vyg}, neutrino masses~\cite{Fukuda_1998,Ahn_2003,Abe_2011}, and the matter–antimatter asymmetry~\cite{Davidson_2008}. Among the various new physics models, we focus on the neutral gauge interactions including the $Z'$ boson and ALP.
	
The new neutral gauge boson  $Z'$ arises generically in Grand Unified Theories (GUTs), in additional $U(1)_X$ gauge extensions, and in Left--Right Symmetric Models (LRSMs), among other well-motivated new physics constructions. At hadron colliders, extensive searches have been performed in the Drell--Yan process ($pp \to Z'$), in vector boson fusion ($pp\to jj Z'$), and in associated production processes ($pp \to W^\pm Z'$, $pp \to Z Z'$, $pp \to H Z'$, $pp \to \tau Z'$, $pp \to \gamma Z'$, among others)~\cite{ATLAS:2024qvg,CMS:2026fjw}. Various decay channels of the $Z'$ boson have also been investigated in the literature, including $Z' \to \ell^+ \ell^-$~\cite{CMS:2021ctt}, $Z' \to q\bar{q}$ (dijet)~\cite{CMS:2022eud}, $Z' \to \nu\bar{\nu}$ (invisible), $Z' \to Z H$~\cite{CMS:2024hbn,Das:2020rsr,Li:2013ava}, $Z' \to Z \gamma$~\cite{ATLAS:2023kcu}, and $Z' \to W^+W^-$~\cite{CMS:2022pjv}. There are also extensive phenomenological studies of the $Z'$ at other colliders. Prospects at future $e^+e^-$ colliders such as the ILC, CLIC, CEPC and FCC-ee have been studied in Refs.~\cite{Yin:2021rlr,Basso:2009hf,Almeida:2004hj}, while the reach of a high energy muon collider has been investigated in Refs.~\cite{Lu:2023jlr,Lu:2024qrf,Barik:2024kwv}. At these lepton colliders, the accessible channels include resonant production $e^+e^-/\mu^+\mu^-\to Z'\to f\bar f$, and virtual-exchange induced  processes $e^+e^-/\mu^+\mu^-\to f\bar f$. Additional on-shell production modes comprise associated production $e^+e^-/\mu^+\mu^-\to Z'\gamma,\, Z'Z,\, Z'H$, as well as vector boson fusion production $W^+W^-\to Z'$. In addition, $Z'$ production at future electron--proton ($ep$) colliders, in particular the LHeC and the FCC-eh, where the $Z'$ can be produced via $e^-p\to e^-j Z'$~\cite{Rizzo:2008fq}. In this paper, we will focus on the detailed properties of the $Z'WW$ interaction. 
	
In the framework of Extended Gauge Models, the origin of the $Z'WW$ trilinear coupling is well motivated: since the $Z'$ and $W$ bosons are gauge bosons of different gauge symmetry factors, such as $U(1)_X$ and $SU(2)_L$, the tree-level $Z'WW$ vertex is strictly forbidden by gauge symmetry. This coupling can be induced after the spontaneous breaking of the electroweak symmetry, through the mixing of gauge eigenstates. In the $U(1)_X$ extension model adopted in this work, the trilinear $Z'WW$ coupling arises dominantly from the $Z$–$Z^\prime$ mixing, with its strength governed by both the new gauge coupling and the mixing angle. Since the coupling structure of the $Z^\prime$ boson is strongly dependent on the gauge structure of the underlying theoretical model, its interaction characteristics can directly reflect the structure of the beyond the SM (BSM) gauge group, boson mixing mechanism and electroweak symmetry breaking pattern, rendering $Z^\prime$ a powerful probe for testing various new physics scenarios~\cite{Zhao:2024fgc}.
	
Beyond the neutral gauge boson discussed above, there exists another class of special new pseudoscalar particles, namely axion. The paradigmatic example is the quantum chromodynamics (QCD) axion, introduced to resolve the strong CP problem in QCD~\cite{Dine2000,KimCarosi2010} through the PQ mechanism~\cite{PecceiQuinn1977a,PecceiQuinn1977b}. A global $U(1)$ symmetry is spontaneously broken at a high energy scale, yielding the axion as a pseudo-Nambu--Goldstone boson.
In this original framework, the decay constant $f_a$ serves as the sole parameter governing both the axion mass $M_a$ and all of its couplings. The ALP generalizes this construction by decoupling $M_a$ from the interaction strengths, rendering them independent free parameters. Phenomenologically, the ALP admits a rich variety of decay channels, including $a \to \gamma\gamma$, $a \to Z\gamma$, $a \to W^+W^-$, $a \to ZZ$, $a \to \ell^+\ell^-$, and $a \to q\bar{q}$~\cite{Ding:2024djo,CMS:2021xor}.
	
	Previous experimental results are primarily concentrated on the ALP--photon coupling $g_{a\gamma\gamma}$, and the constraints on this coupling are very stringent over a wide range of ALP masses. As a result, the $a\gamma\gamma$ coupling is expected to be small~\cite{2018JHEP...09..028C,Bauer:2017ris}, which has motivated the study of photophobic ALPs. The photophobic ALP is characterized by a suppressed coupling to photons, while its dominant interactions occur with other electroweak gauge bosons. This aligns well with our goal of investigating the trilinear $aWW$ coupling. 
    It is important to note that, even though the photophobic condition $g_{a\gamma\gamma}=0$ is imposed at tree level, higher-order effects such as radiative corrections and renormalization-group running can regenerate a small effective $a\gamma\gamma$ coupling. In the context of our phenomenological study, these corrections are well below the sensitivity level and can be safely neglected~\cite{Bauer:2017ris,2018JHEP...09..028C,Ding:2024djo}.
    While several previous studies have individually examined the trilinear couplings of the $aWW$ ~\cite{Mao:2024kgx,Ding:2025wvb,Bharucha:2022zqk,Adhikary:2025hjn,ATLAS:2014vka,CMS:2012gfa,Mimasu:2014tqa,Ebadi:2019vvg,Bonilla:2022jqt,Dutta:2023abe,PhysRevD.108.055040} and $Z'WW $ ~\cite{CMS:2017mhc,ATLAS:2019cbc,Accomando:2017tqg}, a systematic comparative analysis of these two couplings at the collider has not yet been performed.  In this paper, we will investigate the properties of the trilinear couplings of $Z'WW$ and $aWW$ at the LHC and distinguish their similarities and differences. 

	The  paper is organized as follows. In Section~\ref{sec:MD}, we introduce the theoretical framework of  models. 
In Section~\ref{sec:RD}, we study the production of  $Z^\prime$ and $a$ at the LHC and  investigate the properties of the trilinear couplings of $Z^\prime WW$ and $aWW$. Finally, we give a short summary in Section~\ref{sec:sum}.

	\section{Theoretical Framework}\label{sec:MD}
	In this work we consider two representative BSM scenarios: one is the $Z'$ boson from a $U(1)_X$ extension of the SM, and another is the photophobic ALP described within an effective field theory (EFT) framework.
	\subsection{The $U(1)_X$ Model}
	One simple extension of the SM based on the gauge group $SU(3)_C\otimes SU(2)_L\otimes U(1)_Y\otimes U(1)_X$, where $U(1)_X$ is realized as a linear combination of the SM $U(1)_Y$ and the $U(1)_{B-L}$ symmetries~\cite{Mohapatra:1980qe}, known as the $U(1)_X$ extension of the SM~\cite{Das:2021esm}. 
	In addition to the SM fields, the model includes three generations of right-handed neutrinos (RHNs) to cancel all gauge and mixed gauge-gravity anomalies~\cite{PhysRevD.92.015026,PhysRevD.93.115038}, and a SM-singlet scalar $\Phi$ that generates the Majorana mass terms for the RHNs upon the spontaneous breaking of the $U(1)_X$ gauge symmetry.
	The full particle contents of the theory are summarized in TABLE.\ref{tab:U1}~\cite{Chiang:2019ajm,Okada:2016tci}.
	\begin{table}[H]
		\centering
		\begin{tabular}{|c|c|c|c|c|c|c|c|c|}
			\hline
			Gauge group & $q_L^i$ & $u_R^i$ & $d_R^i$ & $\ell_L^i$ & $e_R^i$ & $N_R^i$ & $H$ & $\Phi$ \\
			\hline
			$SU(3)_C$ & $\mathbf{3}$ & $\mathbf{3}$ & $\mathbf{3}$ & $\mathbf{1}$ & $\mathbf{1}$ & $\mathbf{1}$ & $\mathbf{1}$ & $\mathbf{1}$ \\
			\hline
			$SU(2)_L$ & $\mathbf{2}$ & $\mathbf{1}$ & $\mathbf{1}$ & $\mathbf{2}$ & $\mathbf{1}$ & $\mathbf{1}$ & $\mathbf{2}$ & $\mathbf{1}$ \\
			\hline
			$U(1)_Y$ & $1/6$ & $2/3$ & $-1/3$ & $-1/2$ & $-1$ & $0$ & $1/2$ & $0$ \\
			\hline
			$U(1)_X$  & $\frac{1}{6}x_H + \frac{1}{3}x_\Phi$ & $\frac{2}{3}x_H + \frac{1}{3}x_\Phi$ & $-\frac{1}{3}x_H + \frac{1}{3}x_\Phi$ & $-\frac{1}{2}x_H - x_\Phi$ & $-x_H - x_\Phi$ & $-x_\Phi$ & $-\frac{x_H}{2}$ & $2x_\Phi$ \\
			\hline
		\end{tabular}
		\renewcommand{\thetable}{\Roman{table}}
		\caption{Particle contents of the $U(1)_X$ model where $(i = 1, 2, 3)$ represents the family index. The scalar charges $x_H$, $x_\Phi$ are real parameters. The $B - L$ case is obtained with the choice $x_H = 0$ and $x_\Phi = 1$.}\label{tab:U1}
	\end{table}
	
	For our subsequent analysis, we choose the benchmark scenario with $x_H = 1$ and $x_\Phi = 1$.
	The Lagrangian of the gauge field in the model can be expressed in the following form
	\[
	\mathcal{L}_{\text{gauge}} = -\frac14 G_{\mu\nu}^a G^{a\mu\nu} -\frac14 W_{\mu\nu}^i W^{i\mu\nu} -\frac14 B_{\mu\nu} B^{\mu\nu} -\frac14 B'_{\mu\nu} B'^{\mu\nu},
	\tag{2.1}
	\label{eq:2.1}
	\]
	where $W_{\mu\nu}^i$ is the field strength tensor of $SU(2)_L$ ($i=1,2,3$), and $B_{\mu\nu}$, $B'_{\mu\nu}$ are the field strengths of $U(1)_Y$ and $U(1)_X$, respectively.
	In the standard electroweak theory, the field strength tensor of $SU(2)_L$ is defined as
	\[
	W_{\mu\nu}^i = \partial_\mu W_\nu^i - \partial_\nu W_\mu^i + g_2 \epsilon^{ijk} W_\mu^j W_\nu^k,
	\tag{2.2}
	\label{eq:2.2}
	\]
	where $g_2$ is the $SU(2)_L$ gauge coupling constant, and $\epsilon^{ijk}$ is the fully antisymmetric structure constant ($i,j,k=1,2,3$). Substituting the field strength tensor into the kinetic energy term $-\frac14 W_{\mu\nu}^i W^{i\mu\nu}$ and expanding, we obtain the self-interaction term of the three gauge bosons
	\[
	\mathcal{L}_{WWW} = -\, g_2 \, \epsilon^{ijk} \, \partial_\mu W^i_\nu W^{j\mu} W^{k\nu} + \dots.
	\tag{2.3}
	\label{eq:2.3}
	\] 
    
	After symmetry breaking, the neutral gauge eigenstates $W^3_\mu$, $B_\mu$, and $B'_\mu$ mix through the mass matrix, forming the physical states: the photon $A_\mu$, the SM $Z$ boson $Z_\mu$, and the new gauge boson $Z'_\mu$. In the model, these states are defined as follows:
	\begin{equation}
		\begin{aligned}
			W^3_\mu &= s_W A_\mu + c_W c_\theta Z_\mu - c_W s_\theta Z'_\mu, \\
			B_\mu   &= c_W A_\mu - s_W c_\theta Z_\mu + s_W s_\theta Z'_\mu, \\
			B'_\mu  &= s_\theta Z_\mu + c_\theta Z'_\mu,
		\end{aligned}
		\tag{2.5}
		\label{eq:2.5}
	\end{equation}
	where $s_W = \sin\theta_W$ and $c_W = \cos\theta_W$ are the sine and cosine of the Weinberg angle, while $s_\theta = \sin\theta_{ZZ'}$ and $c_\theta = \cos\theta_{ZZ'}$ denote the $Z$--$Z'$ mixing angle. The $Z'WW$ interaction term is obtained as
	\[
    \mathcal{L}_{Z'WW} = i\, g_2\, c_W\, s_\theta \left( Z'_{\mu\nu} W^{+\mu} W^{-\nu} + \left( W^+_{\mu\nu} W^{-\mu} - W^-_{\mu\nu} W^{+\mu} \right) Z'^\nu \right).
    \tag{2.6}
	\label{eq:2.6}
	\]

	\subsection{The Photophobic ALP Effective Field Theory}
	In the EFT description of the leading interactions of the ALP with the SM fields are restricted to the electroweak gauge-field strengths. The effective linear Lagrangian is~\cite{Brivio:2017ije}
	\[
	\mathcal{L}_{\text{eff}} \supset \frac{1}{2} (\partial_\mu a) (\partial^\mu a) - \frac{1}{2} M_a^2 a^2 - c_{\tilde{W}} \frac{a}{f_a} W_{\mu\nu}^b \widetilde{W}^{b,\mu\nu} - c_{\tilde{B}} \frac{a}{f_a} B_{\mu\nu} \widetilde{B}^{\mu\nu},
	\tag{2.8}
	\label{eq:2.8}
	\]
	where $W_{\mu\nu}^b$ and $B_{\mu\nu}$ represent the field strength tensors of the $SU(2)_L$ and $U(1)_Y$ gauge groups, and $\widetilde{X}^{\mu\nu} \equiv \frac{1}{2} \epsilon^{\mu\nu\alpha\beta} X_{\alpha\beta}$ (with $X = W^b, B$) denotes the corresponding dual tensor.The ALP mass $M_{a}$ and decay constant $f_a$ are treated as independent parameters in this work, and the physical observables depend only on the ratios $c_{\tilde{W}}/f_a$ and $c_{\tilde{B}}/f_a$. The dimensionless parameters $c_{\tilde{W}}$ and $c_{\tilde{B}}$ quantify the strength of the ALP couplings to the electroweak gauge bosons. After electroweak symmetry breaking, the interactions between ALP and gauge bosons are generally expressed as
	
	\begin{equation}
		\begin{aligned}
			\mathcal{L}_{\text{eff}} \supset & -\frac{1}{4} g_{a \gamma \gamma} a F_{\mu \nu} \widetilde{F}^{\mu \nu}-\frac{1}{2} g_{a \gamma Z} a Z_{\mu \nu} \widetilde{F}^{\mu \nu} \\
			& -\frac{1}{4} g_{a Z Z} a Z_{\mu \nu} \widetilde{Z}^{\mu \nu}-\frac{1}{2} g_{a W W} a W_{\mu \nu}^{+} \widetilde{W}^{- \mu \nu},
		\end{aligned}
		\tag{2.9}
		\label{eq:2.9}
	\end{equation}
	
	here $F_{\mu\nu}$ represents the electromagnetic field strength tensor, while $Z_{\mu\nu}$ and $W_{\mu\nu}^+$ stand for those of the $Z$ and $W$ bosons, respectively. The corresponding coupling constants are expressed as:
	\begin{equation}
		\begin{aligned}
			g_{a\gamma\gamma} &= \frac{4}{f_a}\left(s_W^2 c_{\tilde{W}} + c_W^2 c_{\tilde{B}}\right), \\
			g_{aZZ} &= \frac{4}{f_a}\left(c_W^2 c_{\tilde{W}} + s_W^2 c_{\tilde{B}}\right), \\
			g_{a\gamma Z} &= \frac{2\,s_{2W}}{f_a}\left(c_{\tilde{W}} - c_{\tilde{B}}\right), \\
			g_{aWW} &= \frac{4}{f_a}\,c_{\tilde{W}},
		\end{aligned}
		\tag{2.10}
		\label{eq:2.10}
	\end{equation}
	where $c_{\tilde{W}}$ and $c_{\tilde{B}}$ are the same coefficients as those defined in Eq.~(\ref{eq:2.8}), and $s_{2W} = \sin 2\theta_W$.
	In this work, we impose the photophobic ALP coupling condition as suggested in~\cite{Georgi:1986df}:
	\[
	g_{a\gamma\gamma} = \frac{4}{f_a}\big(s_W^2 c_{\tilde{W}} + c_W^2 c_{\tilde{B}}\big) = 0,
	\tag{2.11}
	\label{eq:2.11}
	\]
	this constraint immediately yields $s_W^2 c_{\tilde{W}} + c_W^2 c_{\tilde{B}} = 0$, establishing a fixed proportional relation between $c_{\tilde{W}}$ and $c_{\tilde{B}}$.
	\section{Collider Signatures of $Z'$ and ALP at  LHC}\label{sec:RD}
	The trilinear $Z'W^+W^-$ and $aW^+W^-$ couplings are investigated through three representative processes in this section.
	Firstly, we study the resonant production $pp \to Z' \to W^+W^-$,  and analyze the angular distribution of the muon pair in the leptonic final state $W^+W^-\to\mu^+\nu_\mu\,\mu^-\bar{\nu}_\mu$. Secondly, we study the associated production of $W^+ Z'$ with the $Z'$ decay into a $W^+W^-$ pair. To distinguish the properties of vector boson and pseudoscalar, we replace the $Z'$ with the photophobic ALP in the same associated production topology, yielding final states identical to those of the preceding channel. By systematically comparing the production cross sections  and the angular distributions of the final-state muons, we employ these observables as a discriminator between the two classes of new particles.
	Then, we extend the same framework to analyze the associated production of a $Z$ boson with the $Z'$ or the ALP.
	\subsection{Resonant production: $Z' \to W^+W^-$}
\begin{figure}[ht]
		\centering
		\begin{minipage}[H]{0.45\linewidth}
			\includegraphics[width=1\textwidth]{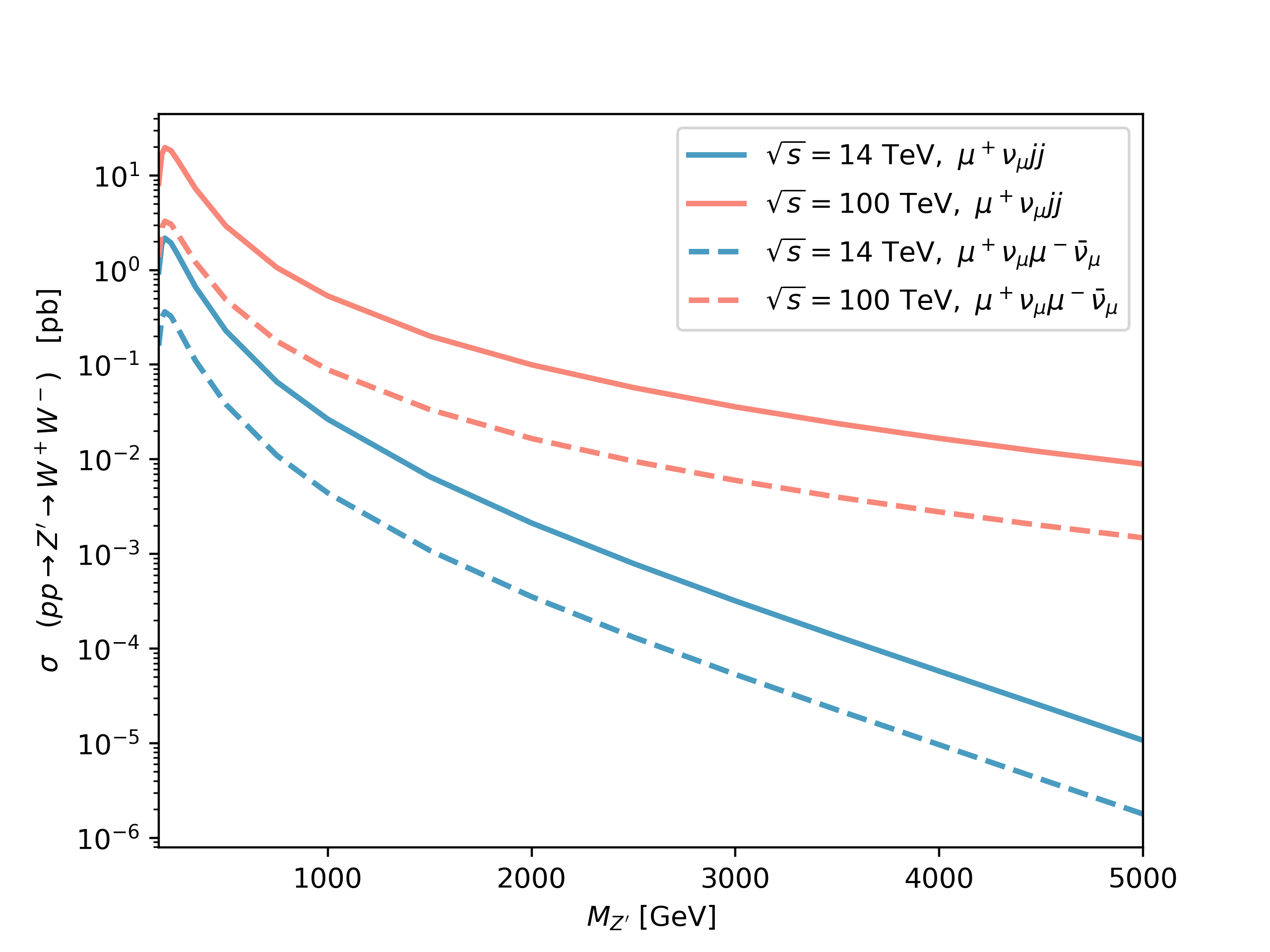}
		\end{minipage}
		\caption{The cross-section distribution versus the $Z'$ boson mass for the $pp\to Z'\to W^+W^-$ process at $\sqrt{s}=14$ and $100\ \text{TeV}$, with the $W$ boson decaying semi-leptonically and fully leptonically.}
		\label{fig:my_label1}
	\end{figure}

    The resonance production of $Z'$ is the most direct process for studying the $Z'$ at $pp$ colliders.
	We focus on the $pp\to Z'\to W^+W^-$ channel, based on the design parameters of hadron colliders such as the HL-LHC at $\sqrt{s} = 14$~TeV and FCC-hh at $\sqrt{s} = 100$~TeV. 
	For final-state reconstruction, we adopt both the semileptonic and the fully leptonic decay modes.
	In the semileptonic mode, the $W^+$ decays to $\mu^+\nu_\mu$, providing a clean trigger signature, while the other $W^-$ boson decays to a dijet, exploiting its larger branching fraction to increase the event yield. 
	This semileptonic configuration thereby combines signal cleanliness with sizeable event rates in the experimental analysis~\cite{ATLAS:2014fik}.
	In the fully leptonic mode, where both $W$ bosons decay into leptons, complementing the analysis with a cleaner final state at the price of a smaller branching fraction.
	The cross sections in these two channels are shown in FIG .~\ref{fig:my_label1} as functions of $M_ {Z'}$ at the two collider energies.
    
	FIG.~\ref{fig:my_label1} displays the production cross section as a function of the $Z'$ boson mass  at $\sqrt{s} = 14\ \text{TeV}$ and $\sqrt{s} = 100\ \text{TeV}$, where the solid and dashed curves correspond to the $W$ boson semileptonic and fully leptonic decaying channels, respectively.
    The mass scan covers from $170$~GeV to $5$~TeV. For the low-mass region, a resonance peak appears near the $W$-pair mass threshold.  The cross section is 2.19~pb at $\sqrt{s} = 14\ \text{TeV}$ and 19.9~pb at $\sqrt{s} = 100\ \text{TeV}$ at resonance peak in the semileptonic channel. With increasing $M_{Z'}$ the production threshold is significantly raised, and the off-shell $s$-channel propagator strongly suppresses the production cross section~\cite{Han:2005mu}. The suppression is further amplified by the weakening of the mixing-induced $Z'WW$ coupling at large $M_{Z'}$. Consequently, the cross section at 14 TeV drops by about five orders of magnitude from the peak to $M_{Z'} = 5$~TeV, while the decline at 100 TeV is considerably milder.
    Although the production falls with mass, the cross section from raising the collider energy grows rapidly in the high-mass region. At small $M_{Z'}$ the cross sections at 100 TeV exceed those at 14 TeV by about one order of magnitude, while the enhancement grows to between one and three orders of magnitude in the medium- and high-mass region. 
     With an integrated luminosity of $20~\mathrm{ab}^{-1}$, a $100$~TeV collider yields $1.8\times10^5$ events in the semileptonic channel and $3\times10^4$ events in the fully leptonic channel for $Z'\to W^+W^-$ at $M_{Z'} = 5$~TeV before selection cuts, whereas a 14 TeV collider with $3~\mathrm{ab}^{-1}$ yields about 30 events in the semileptonic channel at $M_{Z'} = 5$~TeV. The $W^+W^-$ channel may therefore push the sensitivity to heavy $Z'$ bosons into the multi-TeV region at a future $100$~TeV collider, and even the fully leptonic mode provides a clean and usable signal sample despite its smaller branching fraction.
	
\begin{figure}[ht]
		\centering
		\begin{minipage}[H]{0.45\linewidth}
			\includegraphics[width=1\textwidth]{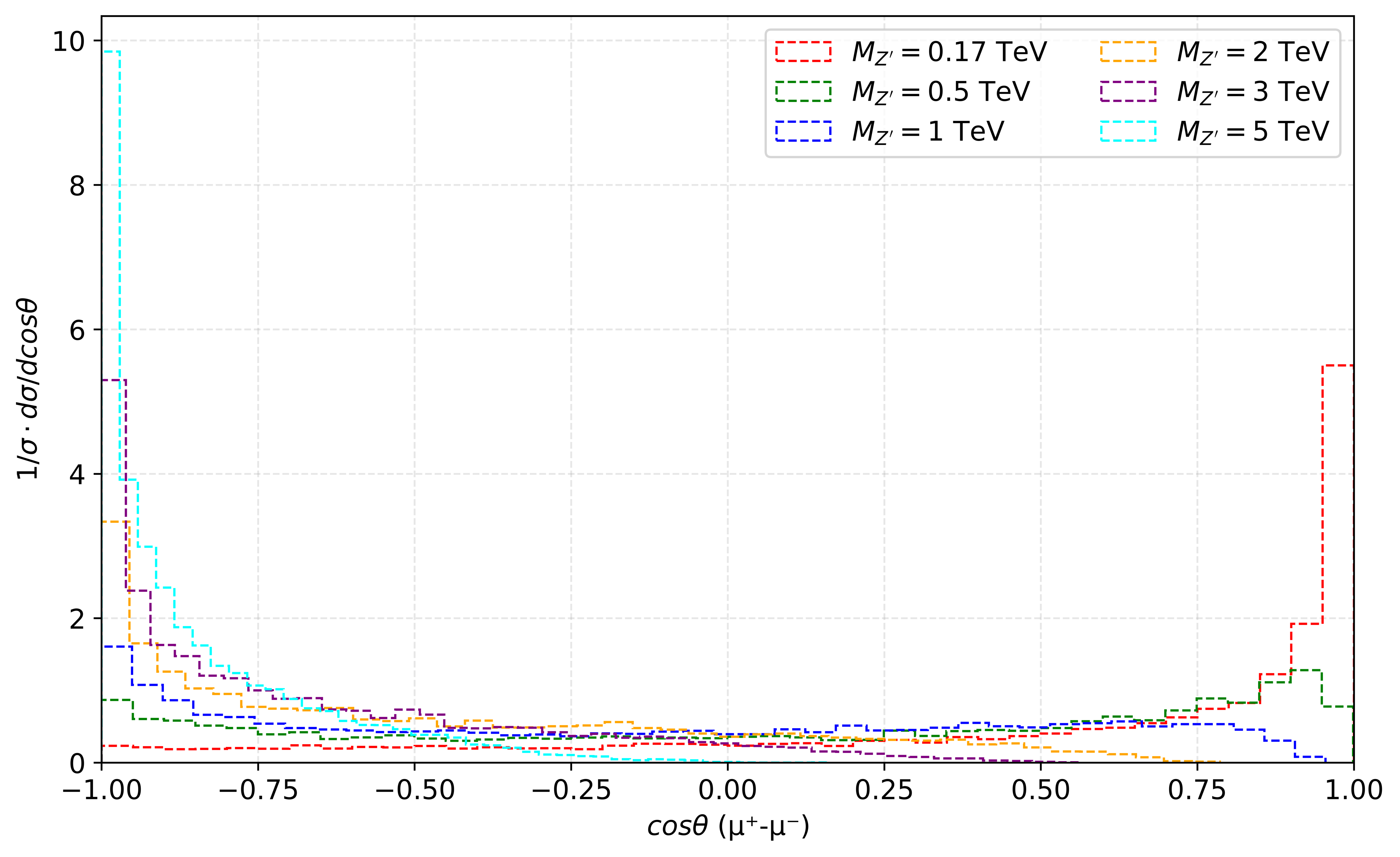}
		\end{minipage}
		\caption{Angular distributions of the final-state $\mu^+\mu^-$ pair from the process $pp \to \mu^+ \mu^- \nu_\mu \bar\nu_\mu$ for different values of $M_{Z'}$ at $\sqrt{s}=14\ \text{TeV}$.}
		\label{fig:my_label2}
	\end{figure}

	FIG.~\ref{fig:my_label2} presents the normalized angular distribution of $\mu^+\mu^-$ pairs in the $pp \to Z' \to W^+ W^- \to \mu^+\mu^-\nu_\mu\bar{\nu}_\mu$ process.  
    The angular distribution is shown as a function of $\cos\theta$,
    \begin{equation}
    \cos\theta = \dfrac{\vec{p}_{\mu_1} \cdot \vec{p}_{\mu_2}}{|\vec{p}_{\mu_1}|\,|\vec{p}_{\mu_2}|},
    \tag{3.1}
	\label{eq:3.1}
    \end{equation}
    where $\vec{p}_{\mu_1}$ and $\vec{p}_{\mu_2}$ are the three-momenta of the two muons, and $\cos\theta \ge 0$ and $\cos\theta < 0$ correspond to the forward and backward regions, respectively.
	The angular distribution results indicate  that the $Z'$ resonance with small $M_{Z'}$ is highly boosted and the final-state muons are collimated along the beam direction. Thus the distribution exhibits a pronounced forward enhancement; the smaller the mass, the stronger the boost effect, resulting in a larger forward peak. To quantify the forward--backward difference more directly, we define the forward--backward asymmetry as
    \begin{equation}
    A_{FB} = \frac{\sigma(\cos\theta \ge 0) - \sigma(\cos\theta < 0)}{\sigma(\cos\theta \ge 0) + \sigma(\cos\theta < 0)}~.
    \tag{3.2}
	\label{eq:3.2}
    \end{equation}
    Across the interval $M_{Z'} = 0.17$--$5$~TeV, the asymmetry $A_{FB}$ decreases monotonically from positive to negative values. For instance, $A_{FB}$ = 0.57 at $M_{Z'} = 0.17$~TeV, $A_{FB}$ = 0.19 at $M_{Z'} = 0.5$~TeV, and $A_{FB} = -0.72$ at $M_{Z'} = 2$~TeV. This trend reflects the fact that the Lorentz boost effects of the $Z'$ boson is significantly weakened with large masses and the $W^+W^-$ pair is produced with a wider opening angle, so that the angular distribution of the final-state muons increasingly develops a backward enhancement.
    At $M_{Z'}= 5$~TeV, the forward--backward asymmetry is close to $-1$.

	\subsection{Associated production:  $W^+ Z'/ a~ \to W^+W^+W^-$}
	Having established the baseline features of resonant $Z'$ production, we now turn to a detailed analysis of the associated production process to better understand the trilinear gauge couplings involving the new particles with the $W$ bosons.
    The study of the associated channel also motivates a direct comparison between two distinct new particles, the neutral $Z'$ and the ALP.
	\begin{figure}[ht]
		\centering
		\captionsetup[subfigure]{justification=centering}
		\begin{subfigure}[b]{0.45\textwidth}
			\centering
			\includegraphics[width=\linewidth]{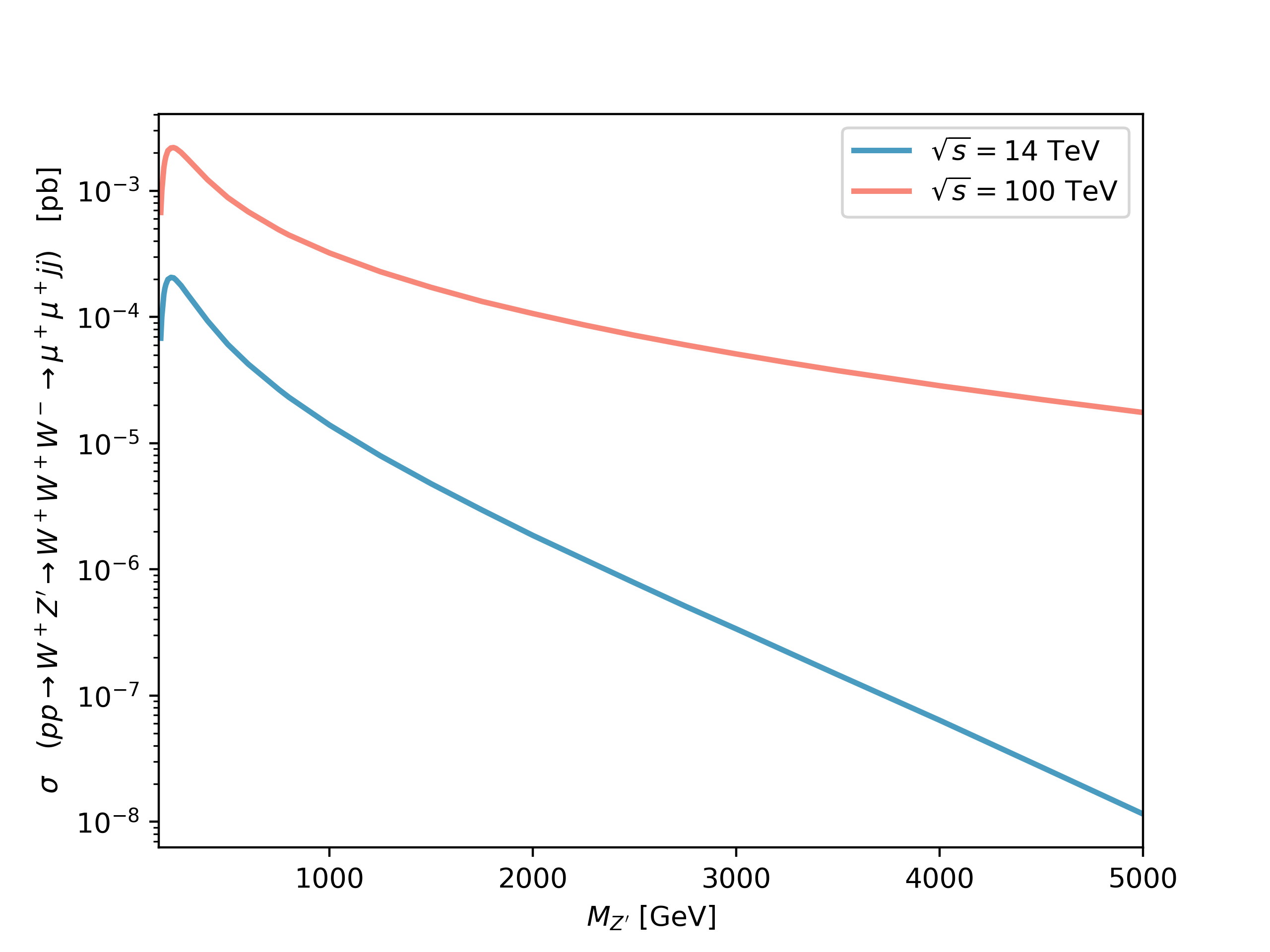}
			\subcaption{}   
			\label{fig:sub-a3}
		\end{subfigure}
		\hfill
		\begin{subfigure}[b]{0.45\textwidth}
			\centering
			\includegraphics[width=\linewidth]{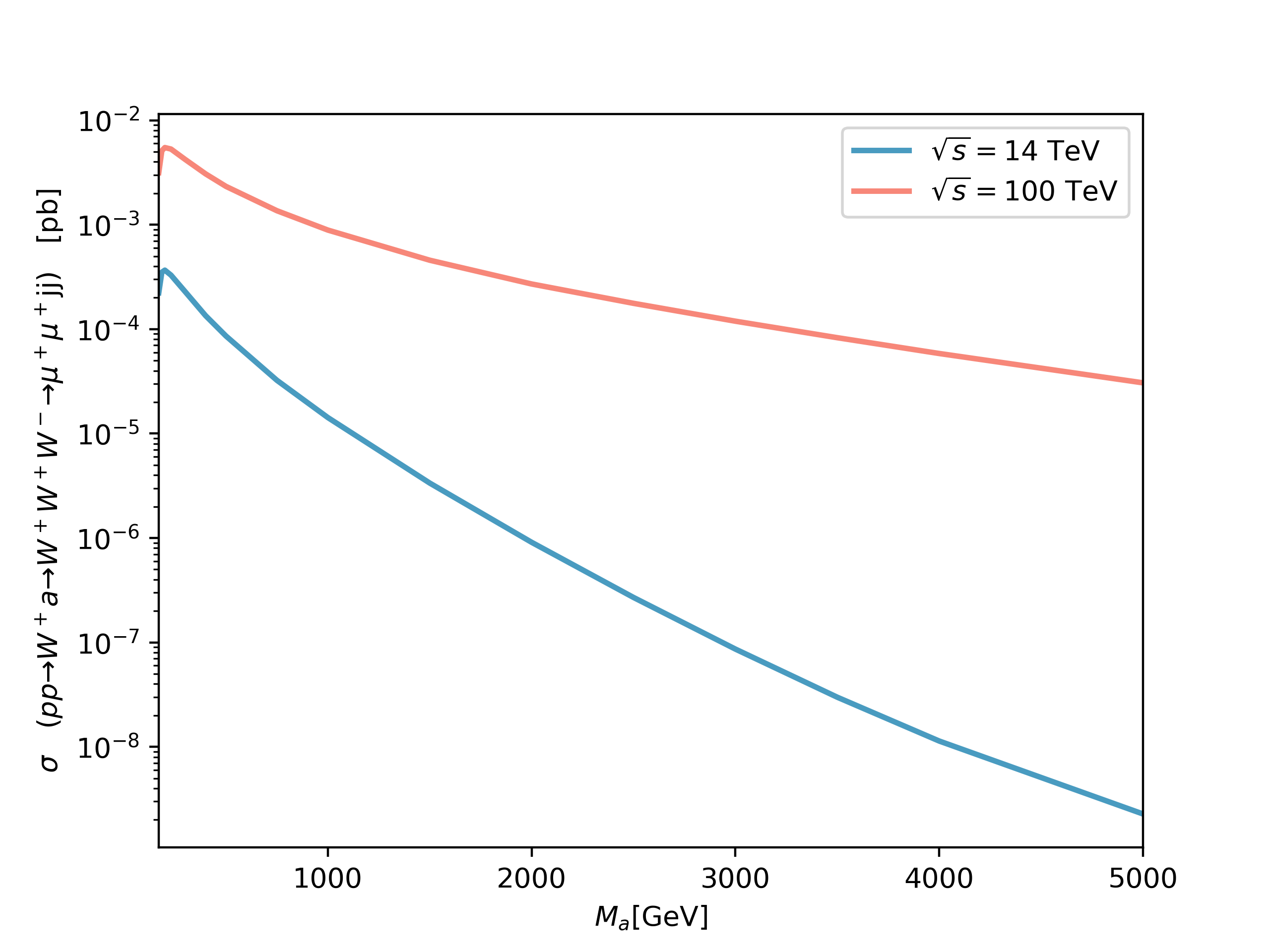}
			\subcaption{}   
			\label{fig:sub-b3}
		\end{subfigure}
		\caption{The cross-section distribution versus the $Z'$ boson mass (a) or the ALP mass (b) for the process $pp \to W^+ Z'/a \to W^+ W^+ W^- \to \mu^+ \mu^+ \nu_\mu \nu_\mu  jj$ at $\sqrt{s}=14$ and $100~\mathrm{TeV}$.}
		\label{fig:my_label3}
	\end{figure}
	
	FIG.~\ref{fig:my_label3} shows the production cross sections at $\sqrt{s} = 14\ \text{TeV}$ and $\sqrt{s} = 100\ \text{TeV}$ for the two associated production channels $pp\to W^+Z'\to W^+W^+W^-\to\mu^+\mu^+\nu_\mu\nu_\mu  jj$ in FIG~\ref{fig:my_label3}(\subref{fig:sub-a3}) and $pp \to W^+ a \to W^+ W^+ W^-\to\mu^+\mu^+\nu_\mu\nu_\mu  jj$ in FIG~\ref{fig:my_label3}(\subref{fig:sub-b3}).
	In both channels, the cross section first rises to a pronounced peak at small values of $M_{Z'}$ or $M_{a}$ and then decreases monotonically as the corresponding mass increases. With the ALP-associated cross section falling off by about 5 orders of magnitude and the $Z'$ - associated one by about 3 orders of magnitude across the scanned mass range at $\sqrt{s} = 14\ \text{TeV}$, the plots reflect the different mass dependence from the $aWW$ and $Z'WW$ couplings. 
	Because of the additional electroweak coupling and the suppression of the final-state phase space, the cross sections of the associated production processes are lower than that of the resonant channel by a factor of about 2-4 in the $Z'$ sector.
    
	Owing to the different spin assignments, namely spin-1 for the $Z'$ and spin-0 for the ALP, and the distinct tensor structures of their couplings, the two particles produce distinct angular correlations among the final-state particles. 
	Consequently, a systematic comparison of the muon angular distributions serves as an effective discriminator between the two classes of new particles. 
    To better compare the coupling properties of the $Z'$ and the photophobic ALP, we study the angular distributions of the final-state muons within a common final-state topology.
	In the semileptonic final state, muons can originate either from the associated $W^+$ decay or from the $W^+W^-$ pair produced by the new-particle decay, giving rise to same-sign and opposite-sign dimuon final states. 
	In the following, we present and compare the muon angular distributions in the associated production process $pp\to W^+Z'/a$ separately for each case.

	\begin{figure}[ht]
		\centering
		\captionsetup[subfigure]{justification=centering}
		\begin{subfigure}[b]{0.45\textwidth}
			\centering
			\includegraphics[width=\linewidth]{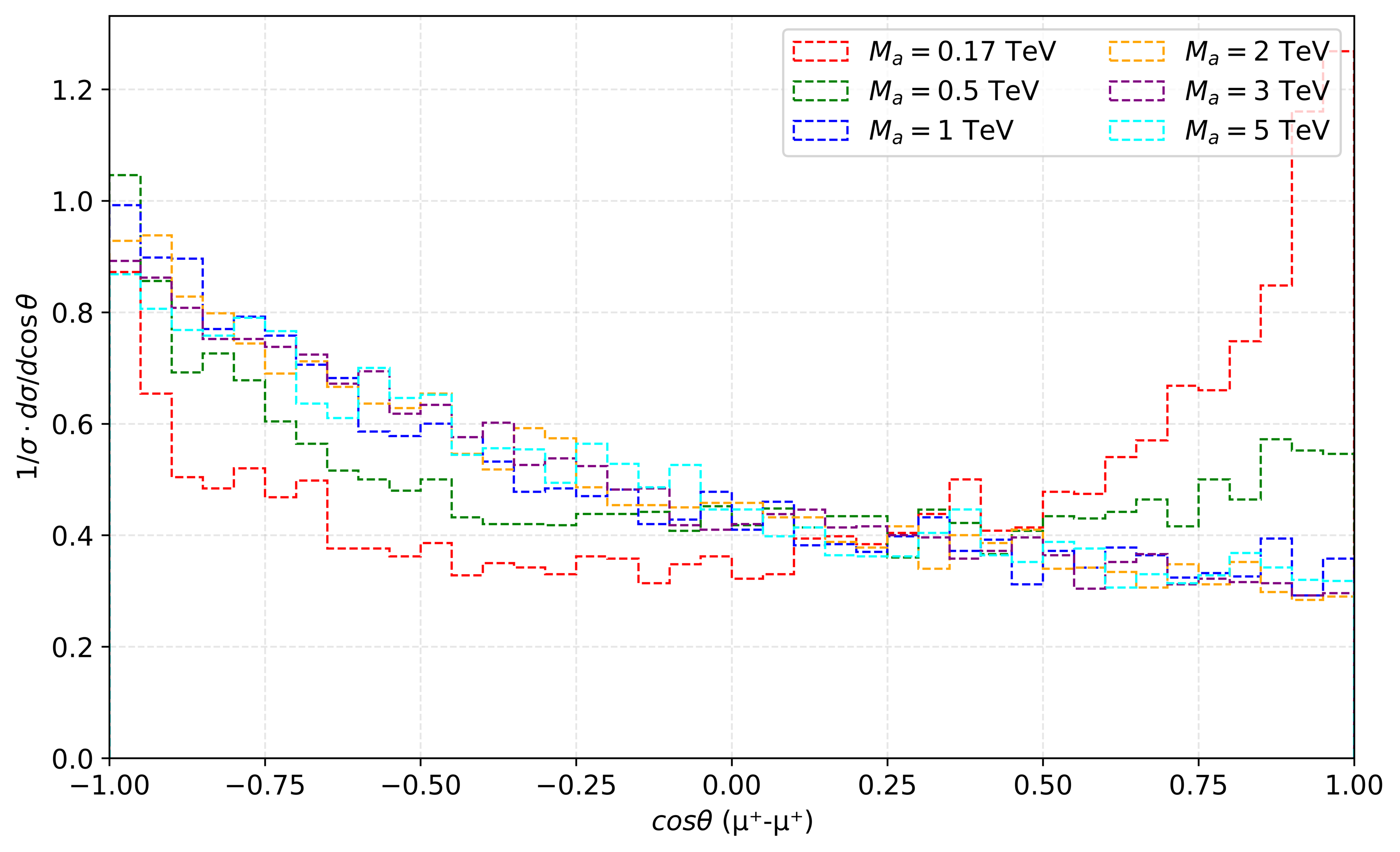}
			\subcaption{}   
			\label{fig:sub-a4}
		\end{subfigure}
		\hfill
		\begin{subfigure}[b]{0.45\textwidth}
			\centering
			\includegraphics[width=\linewidth]{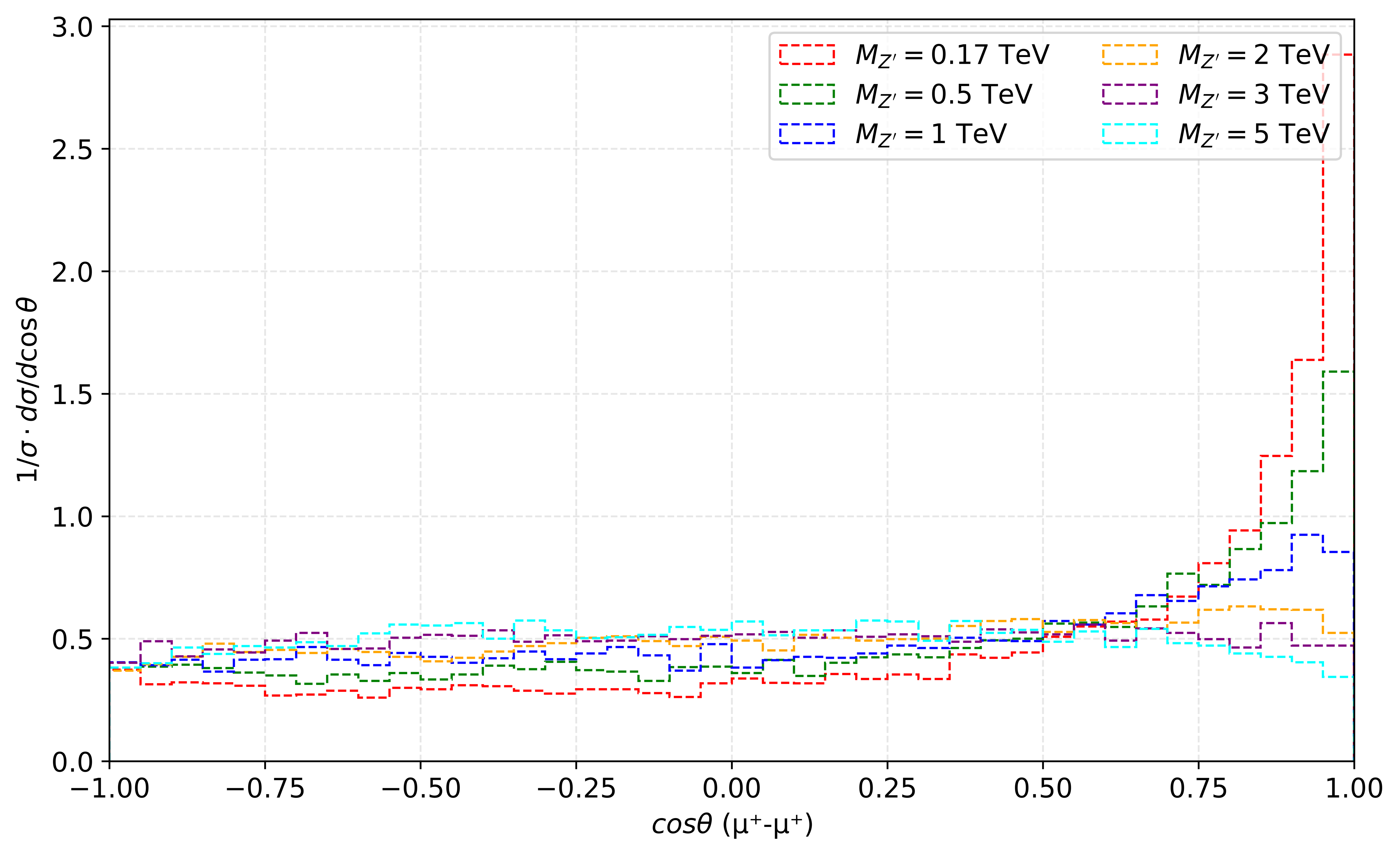}
			\subcaption{}   
			\label{fig:sub-b4}
		\end{subfigure}
		\caption{Angular distributions of the final-state $\mu^+\mu^+$ pair from the process $pp\to W^+Z'/a  \to W^+ W^+\allowbreak  W^-  \to   \mu^+\mu^+\nu_\mu\nu_\mu jj$ for different values of $M_a$ and $M_{Z'}$ at $\sqrt{s} = 14\ \text{TeV}$.}
		\label{fig:my_label4}
	\end{figure}
	FIG.~\ref{fig:my_label4}(\subref{fig:sub-a4}) presents the normalized angular distribution of like-sign $\mu^+\mu^+$ pairs in the process of $pp\to W^+a\to W^+W^+W^-\to\mu^+\mu^+ \nu_\mu\nu_\mu jj$, where the two like-sign muons originate from the decays of two $W^+$ bosons. The angular distribution results indicate that when the mass of the pseudoscalar particle $a$ is less than 500 GeV, the final-state muons tends to be emitted along the beam direction or to exhibit a back-to-back distribution, with the curve showing high values at both ends and a dip in the middle. The smaller the mass, the stronger the boost effect, resulting in a larger forward peak. As the mass of the $a$ particle increases, the Lorentz boost effect is significantly weakened, and the angular distribution of the final-state muons increasingly approaches a backward distribution characteristic.
	FIG.~\ref{fig:my_label4}(\subref{fig:sub-b4}) presents the normalized angular distribution of the final state $\mu^+\mu^+$ from the process of $pp\to W^+Z'\to W^+W^+W^-\to\mu^+\mu^+ \nu_\mu\nu_\mu jj$ . When the $Z'$ mass $M_{Z'} < 2\,\mathrm{TeV}$, the angular distribution exhibits a pronounced peak at $\cos\theta = 1$ due to the boost effect.
	In the high-mass range, the significantly weakened Lorentz boost effect leads to the disappearance of the forward-scattering peak, a flattened angular distribution, and the loss of forward–backward asymmetry. At $M_{Z'} = 5$~TeV, $A_{FB} = 0.001$. This feature stems from the decay dynamics of the two $\mu$ originating from different parent bosons.

	\begin{figure}[ht]
		\centering
		\captionsetup[subfigure]{justification=centering}
		\begin{subfigure}[b]{0.45\textwidth}
			\centering
			\includegraphics[width=1\linewidth]{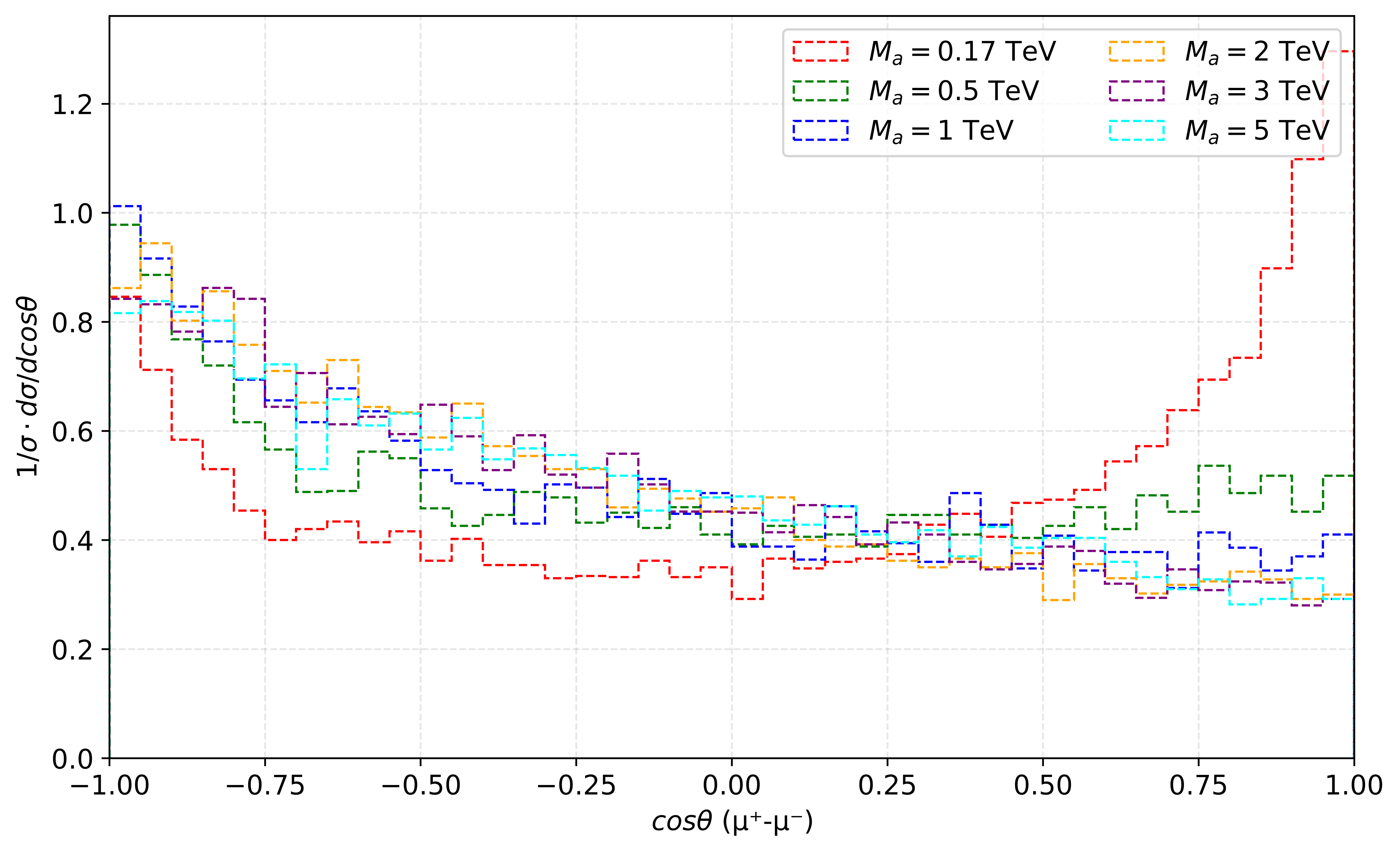}
			\subcaption{}
			\label{fig:sub-a5}
		\end{subfigure}
		\hfill
		\begin{subfigure}[b]{0.45\textwidth}
			\centering
			\includegraphics[width=\linewidth]{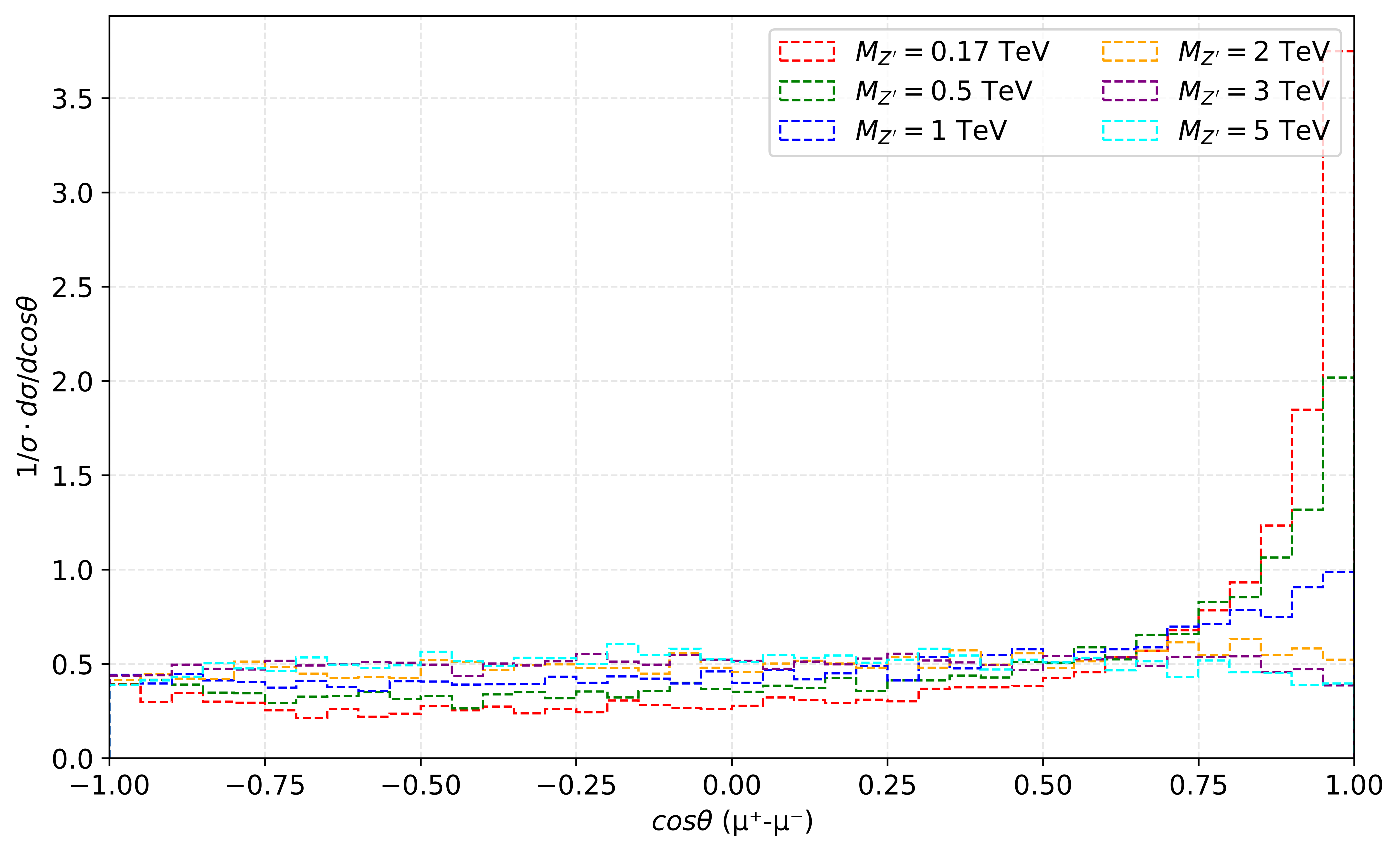}
			\subcaption{}
			\label{fig:sub-b5}
		\end{subfigure}
		\caption{Angular distributions of the final-state $\mu^+\mu^-$ pair from the process $pp\to W^+Z'/a  \to W^+ W^+\allowbreak  W^-\to\mu^+\mu^-\nu_\mu\bar\nu_\mu jj$ for different values of $M_a$ (a) and $M_{Z'}$ (b) at $\sqrt{s}=14\ \text{TeV}$. $\mu^+$ is from the associated $W^+$ boson, and $\mu^-$ from $a/Z'$-decayed $W^-$.}
		\label{fig:my_label5}
	\end{figure}
    
	Building upon the above analysis, we further extend our study to the final state to $\mu^+ \mu^- \nu_\mu \bar{\nu}_\mu jj$, where the $W^-$ boson decays into $\mu^-$, and the associated $W^+$ boson decays into $\mu^+$. 
	FIG.~\ref{fig:my_label5} presents the angular distribution of the $\mu^+\mu^-$ system in this final state.
	The overall shape of this angular distribution resembles that of the aforementioned same-sign muon final states, demonstrating that the muon charge configuration has a negligible impact on the spatial angular distribution of final-state particles.  
	
	\begin{figure}[ht]
		\centering
		\captionsetup[subfigure]{justification=centering}
		
		\begin{subfigure}[b]{0.45\textwidth}
			\centering
			\includegraphics[width=\linewidth]{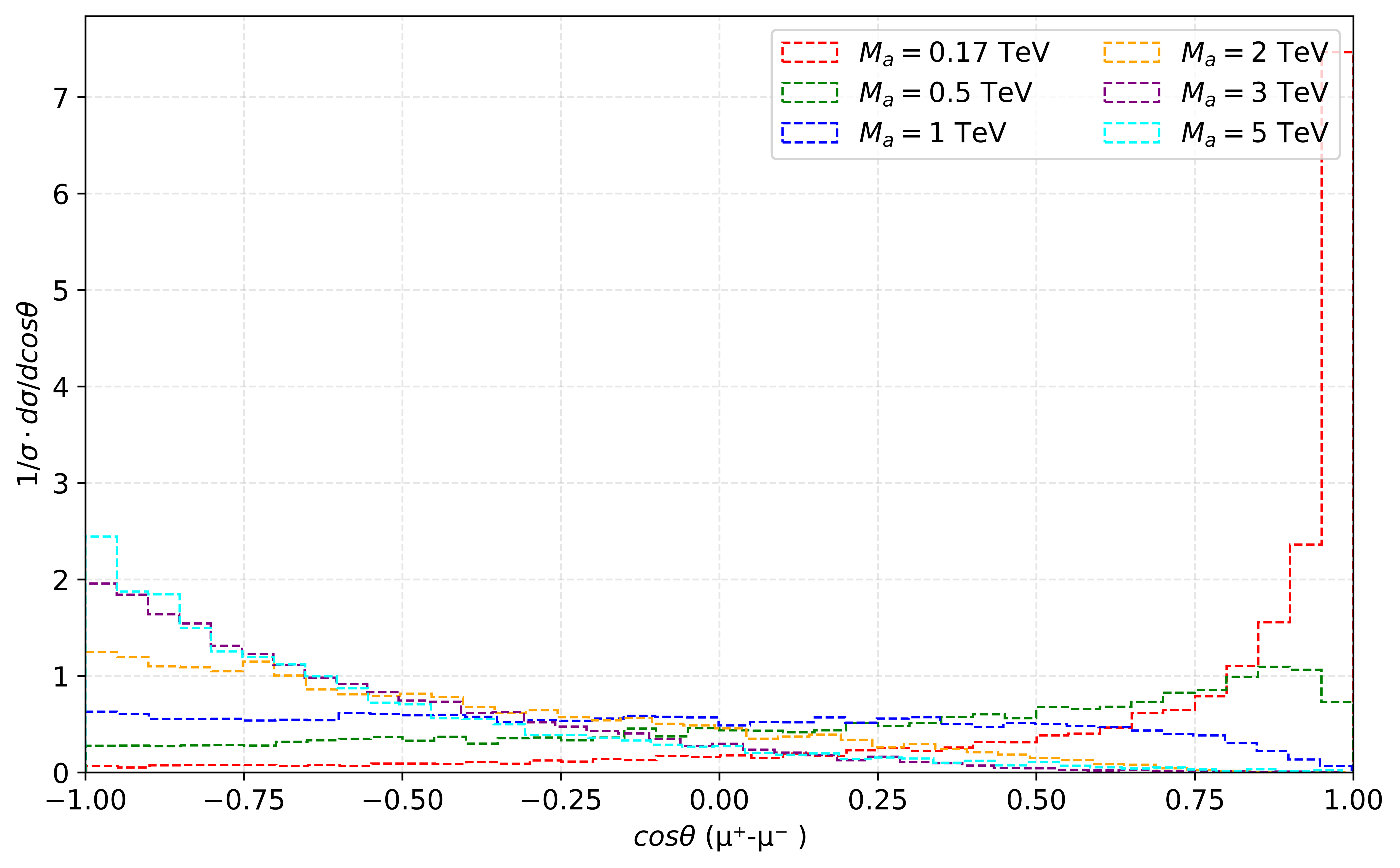}
			\subcaption{}
			\label{fig:sub-a6}
		\end{subfigure}
		\hfill
		\begin{subfigure}[b]{0.45\textwidth}
			\centering
			\includegraphics[width=\linewidth]{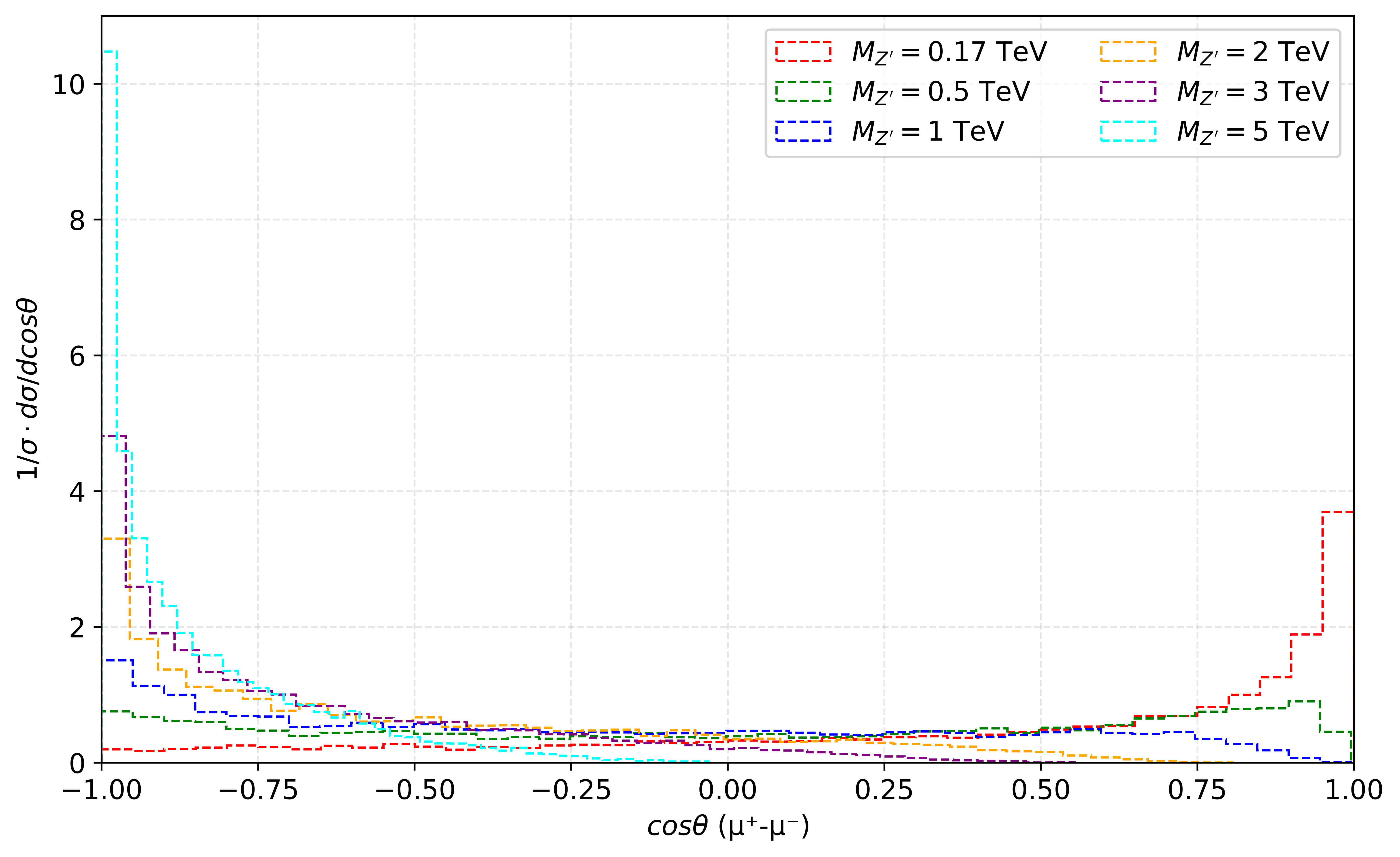}
			\subcaption{}
			\label{fig:sub-b6}
		\end{subfigure}
		\caption{Angular distributions of the final-state $\mu^+\mu^-$ pair from the process $pp\to W^+Z'/a  \to W^+ W^+\allowbreak  W^-\to\mu^+\mu^-\nu_\mu\bar\nu_\mu jj$ for different values of $M_a$ (a) and $M_{Z'}$ (b) at $\sqrt{s}=14\ \text{TeV}$. $\mu^+\mu^-$ are both from $a/Z'$-decayed particles.}
		\label{fig:my_label6}
	\end{figure}
	
	However, when both opposite-sign muons originate from the decay of the new particle, their angular distribution may carry more direct information about the parent particle. The corresponding angular distribution is presented in FIG.~\ref{fig:my_label6}.
	The (\subref{fig:sub-a6}) presents the results for the ALP scenario.
	At low $M_{a}$, the distribution shows a pronounced forward enhancement. The decay products of the highly boosted $a$ are moving in the forward direction, while the backward event yield is strongly suppressed. 
	As $M_{a}$ increases, the shape evolves markedly. Around the transition mass $M_a\approx 0.85\ \mathrm{TeV}$ the forward--backward asymmetry is close to $0$, and for $M_a>0.85\ \mathrm{TeV}$ its sign reverses, the distribution shifting toward a backward enhancement. 
	At $M_a=2\ \mathrm{TeV}$, the contribution from the exactly forward configuration with $\cos\theta=+1$ vanishes.
	With increasing $M_{a}$, the muon-pair distribution shifts toward the backward direction.
	The (\subref{fig:sub-b6}) presents the results for the $Z'$ scenario.
	The angular distributions for the $Z'$ and the ALP show the same qualitative trend, a forward peak at low mass and a backward peak at high mass. They differ quantitatively, however, with the ALP producing a stronger forward peak at low mass and the $Z'$ yielding a sharper backward peak at high mass.
	Hence, the detailed $\cos\theta$ distribution rather than merely the sign of $A_{FB}$ serves as a clean discriminator between the two new physics scenarios. 
    The corresponding $A_{FB}$ values are summarized in TABLE~\ref{tab:U2}.
    
\begin{table}[htbp]
\centering
\setlength{\tabcolsep}{10pt}
\begin{tabular}{|c|c|c|c|c|c|c|c|}
\hline
  \makecell{$M_a$ \\ {[TeV]}}
  & \multicolumn{3}{c|}{$A_{FB}$}
  & \makecell{$M_{Z'}$ \\ {[TeV]}}
  & \multicolumn{3}{c|}{$A_{FB}$} \\[-2.0ex]

\cline{2-4} \cline{6-8}
  & A & B & C & & A & B & C \\
\hline
0.17 &  0.14 &  0.13 &  0.81 & 0.17 &  0.41 &  0.45 &  0.52 \\
0.5  & -0.10 & -0.11 &  0.33 & 0.5  &  0.27 &  0.31 &  0.08 \\
1    & -0.26 & -0.22 & -0.14 & 1    &  0.16 &  0.18 & -0.26 \\
2    & -0.28 & -0.29 & -0.64 & 2    &  0.09 &  0.06 & -0.69 \\
3    & -0.27 & -0.27 & -0.84 & 3    &  0.03 &  0.009 & -0.90 \\
5    & -0.27 & -0.25 & -0.80 & 5    &  0.001 & -0.006 & -1.00 \\
\hline
\end{tabular}
\renewcommand{\thetable}{\Roman{table}}
\caption{The $A_{FB}$ values of the muon pairs from the process $pp \to W^+ Z'/a \to W^+ W^+ W^-$ for different values of $M_a$ and $M_{Z'}$ at $\sqrt{s}=14\ \text{TeV}$, where columns A, B, and C correspond to the processes shown in FIGs.~\ref{fig:my_label4}, \ref{fig:my_label5}, and \ref{fig:my_label6}, respectively.}
\label{tab:U2}
\end{table}
	The $A_{FB}$ values provide a more intuitive illustration of the differences in angular distributions among these processes. Taking the process shown in FIG.~\ref{fig:my_label4} as an example, at the same mass point for the $a$ and $Z'$ scenarios, both the sign and the magnitude of $A_{FB}$ exhibit pronounced differences. In particular, at $M_{Z'/a} = 5$~TeV, the $A_{FB}$ for the $a$ scenario is approximately $-0.27$, while that for the $Z'$ scenario approaches $0$. This demonstrates that the angular distribution serves as an important probe for distinguishing between the two particles.
    
	\subsection{Associated production: $Z Z'/a \to Z W^+W^-$}
	We now extend our analysis to the Z-associated channel $pp\to Z Z'/a\to ZW^+W^-$, thereby providing complementary tests of the $Z'/a$ interactions.
	We follow the same analysis strategy as before, focusing on the production cross sections and the angular distributions.
	\begin{figure}[ht]
		\centering
		\captionsetup[subfigure]{justification=centering}
		\begin{subfigure}[b]{0.45\textwidth}
			\centering
			\includegraphics[width=\linewidth]{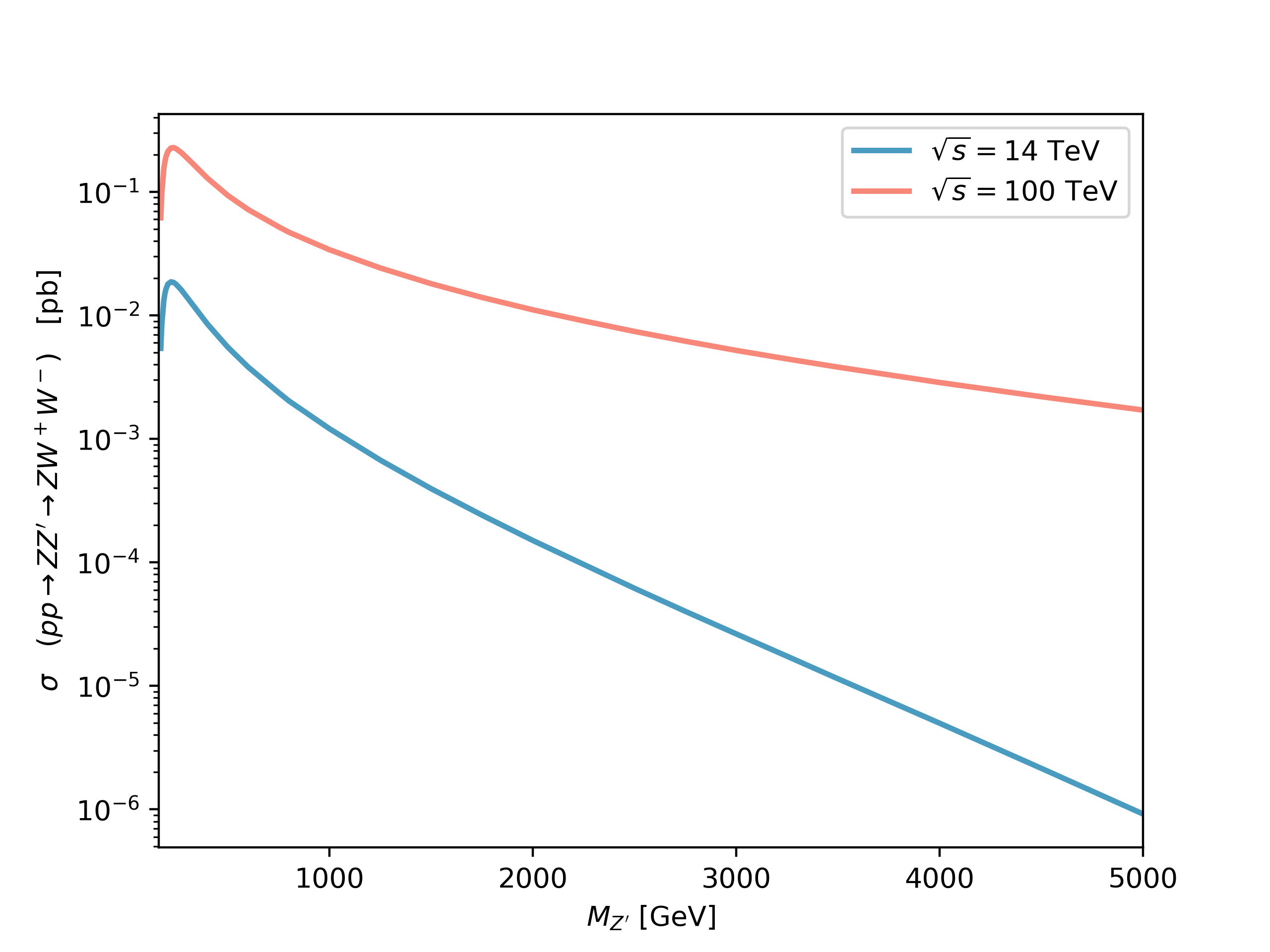}
			\subcaption{}   
			\label{fig:sub-a7}
		\end{subfigure}
		\hfill
		\begin{subfigure}[b]{0.45\textwidth}
			\centering
			\includegraphics[width=\linewidth]{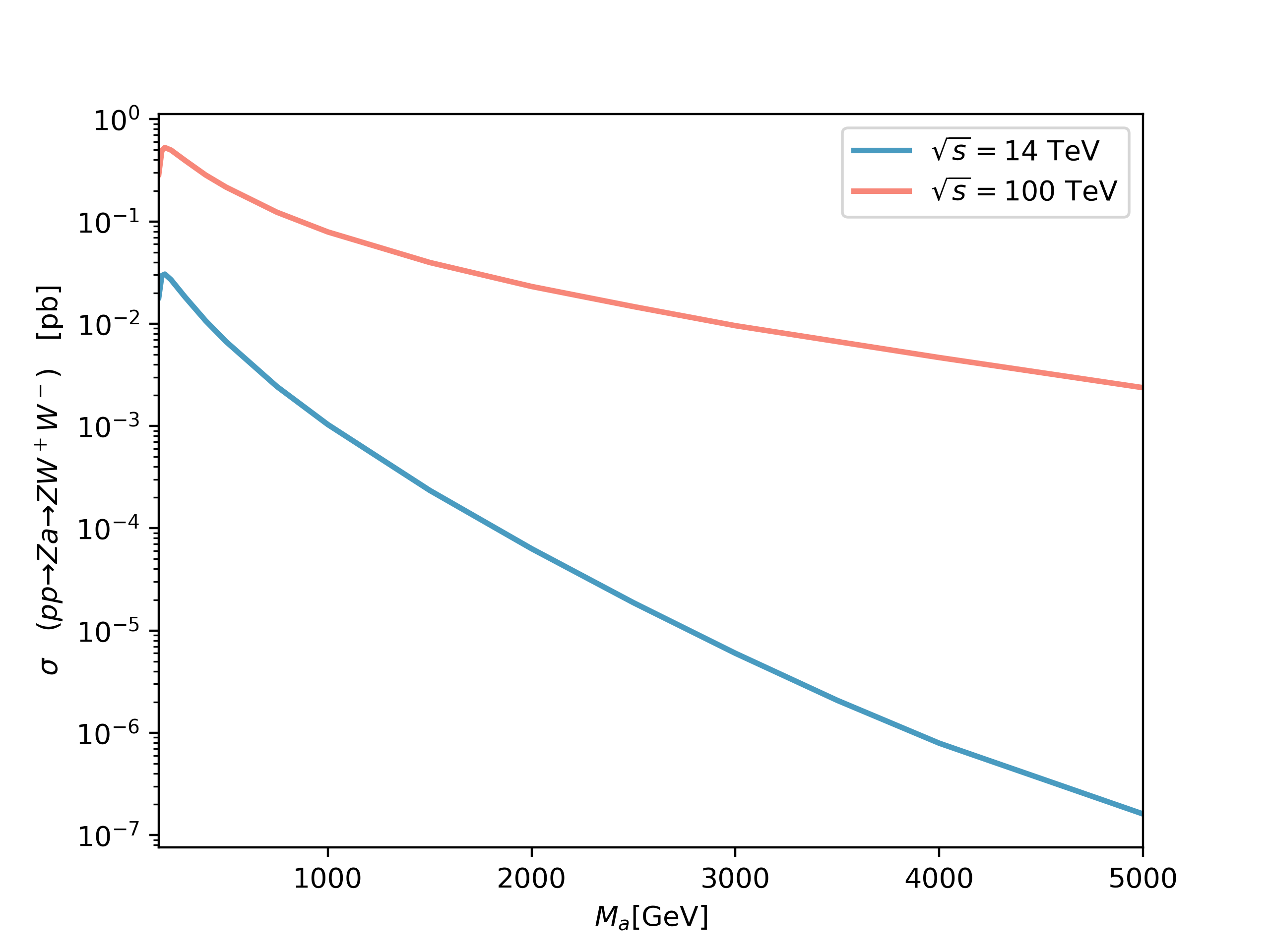}
			\subcaption{}  
			\label{fig:sub-b7}
		\end{subfigure}
		\caption{The cross-section distribution versus the $Z'$ boson mass (a) and the ALP mass (b) for the process $pp\to Z Z'/a\to ZW^+W^-$ at $\sqrt{s}=14$ and $100~\mathrm{TeV}$.}
		\label{fig:my_label7}
	\end{figure}
	FIG.~\ref{fig:my_label7} presents the cross sections for the two associated production processes $pp \to Z Z' \to Z W^+ W^-$ and $pp \to Z a \to Z W^+ W^-$ at $\sqrt{s}=14\ \mathrm{TeV}$ and $\sqrt{s}=100\ \mathrm{TeV}$. 
	The mass dependence of the cross sections for the two channels are qualitatively similar to those observed in the $W^+$ associated case of FIG.~\ref{fig:my_label3}. After peaking at lower masses, the cross section decrease monotonically with mass increases, with the ALP cross section decreasing more steeply than the $Z'$ rate.
    However, a quantitative discrepancy arises when comparing the $W^+$ associated and $Z$ associated production processes decaying to the same final state. For example, for a final state $\mu^+\mu^-\nu_\mu\bar\nu_\mu jj$, where the dimuon originates from the $Z'$, with $M_{Z'} = 3$~TeV and $\sqrt{s}=14\ \mathrm{TeV}$, the cross section for the $W^+$ associated process is $\sigma_{W^+} = 3.4 \times 10^{-7}~\mathrm{pb}$, while that for the $Z$ associated process is $\sigma_{Z} = 1.5 \times 10^{-7}~\mathrm{pb}$. Similarly, when the dimuon originates from the $a$ with $M_{a} = 3$~TeV, yielding $\sigma_{W^+} = 8.7 \times 10^{-8}~\mathrm{pb}$ and $\sigma_{Z} = 3.3 \times 10^{-8}~\mathrm{pb}$, respectively.
    
\begin{figure}[ht]
		\centering
		\captionsetup[subfigure]{justification=centering}
		\begin{subfigure}[b]{0.45\textwidth}
			\centering
			\includegraphics[width=\linewidth]{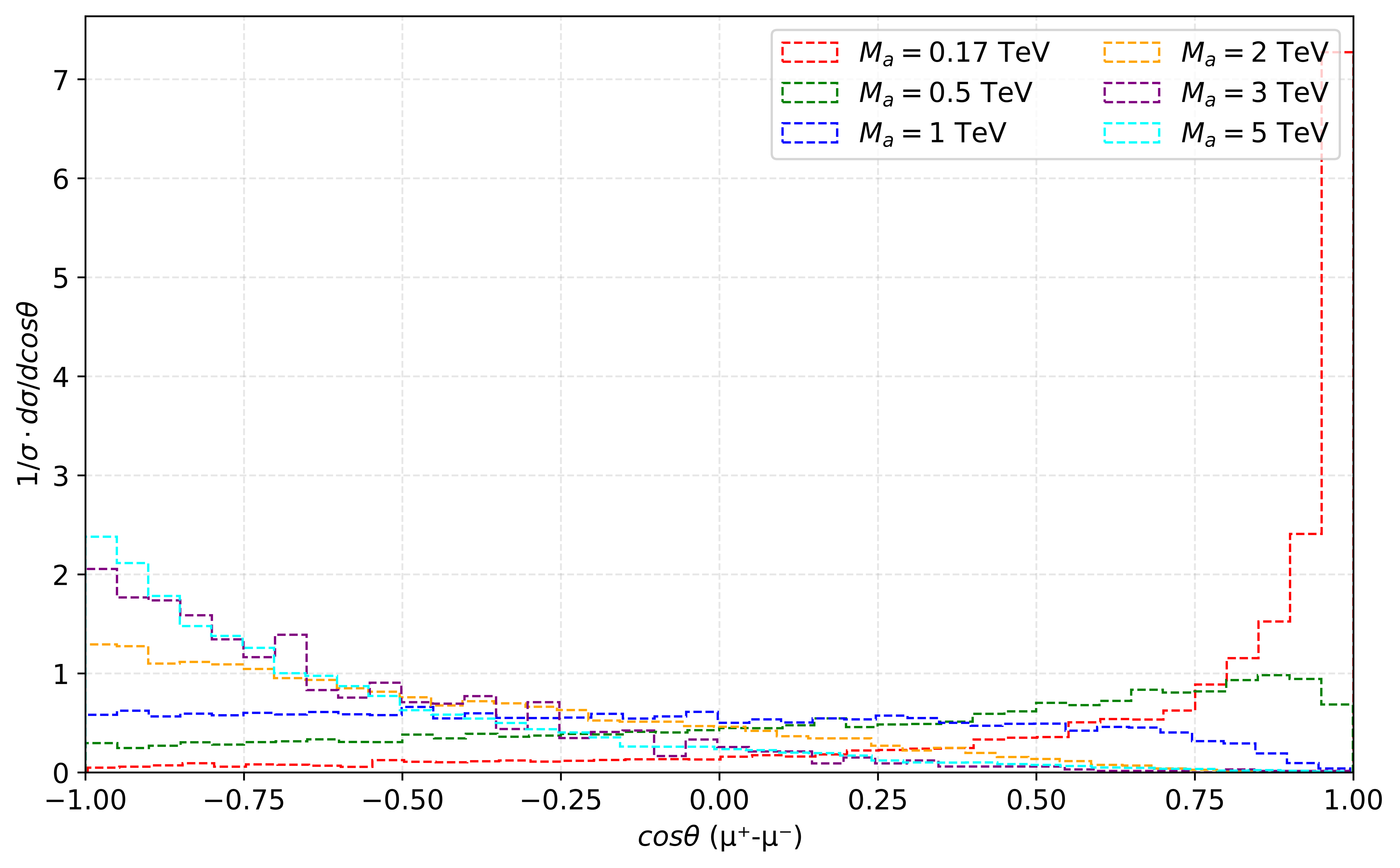}
			\subcaption{}
			\label{fig:sub-a8}
		\end{subfigure}
		\hfill
		\begin{subfigure}[b]{0.45\textwidth}
			\centering
			\includegraphics[width=\linewidth]{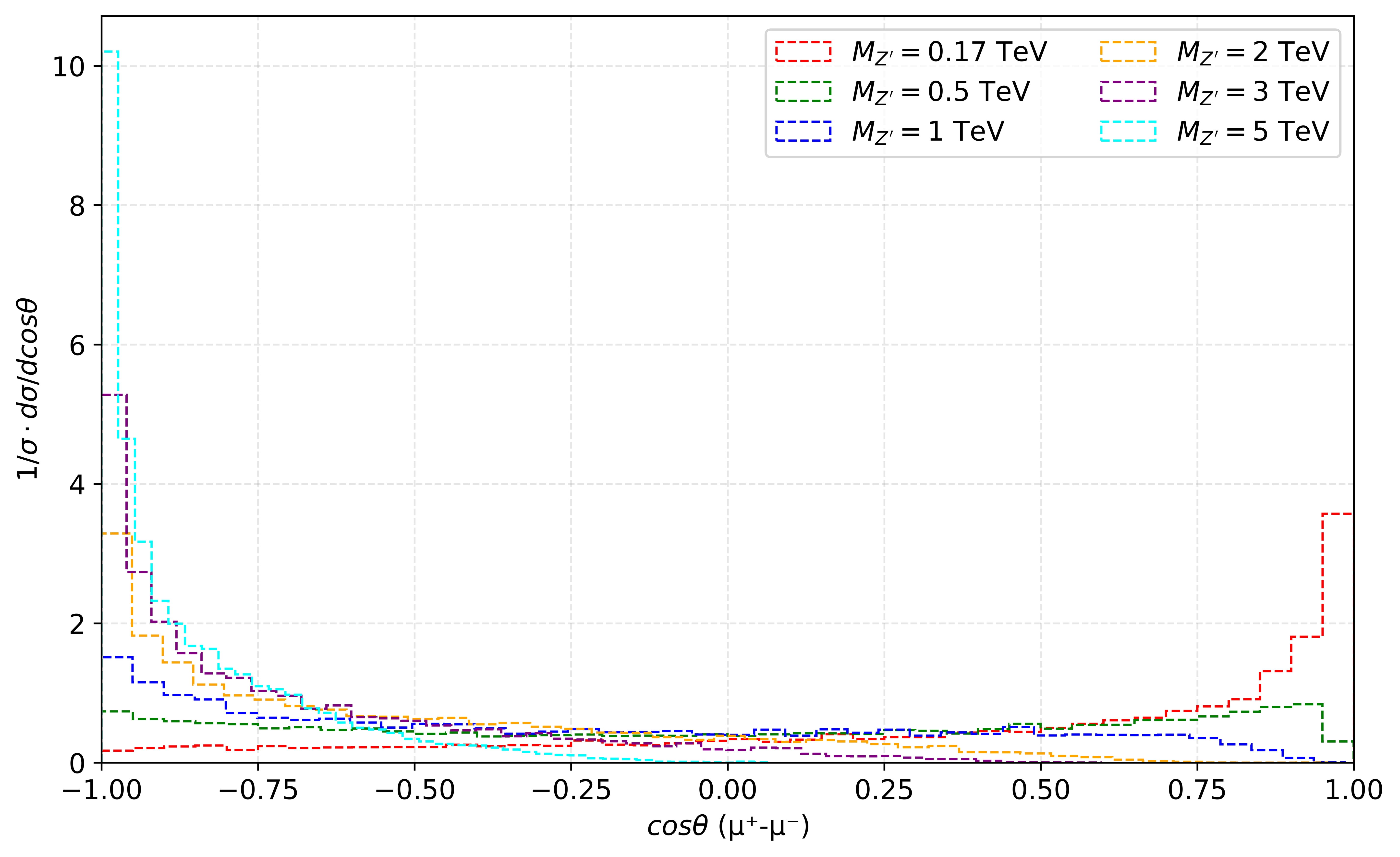}
			\subcaption{}
			\label{fig:sub-b8}
		\end{subfigure}
		\caption{Angular distributions of the final-state $\mu^+\mu^-$ pair from the process $pp\to ZZ'/a\to ZW^+W^-\allowbreak\to\mu^+\mu^-\nu_\mu\bar{\nu}_\mu jj$ for different values of $M_a$ (a) and $M_{Z'}$ (b) at $\sqrt{s}=14\ \text{TeV}$. $\mu^+\mu^-$ is from associated $a/Z'$.}
		\label{fig:my_label8}
	\end{figure}

	In the angular distribution analysis, we separate the events into the semileptonic and fully leptonic channels according to the final-state topology.
    We first focus on the semileptonic channel, for which FIG.~\ref{fig:my_label8} shows the normalized $\mu^+\mu^-$ angular distribution in the associated production process $pp\to Z Z'/a\to ZW^+W^-\to\mu^+\mu^-\nu_\mu\bar\nu_\mu jj$ with the Z decaying into a pair of jets. 
    Both panels reproduce the corresponding distributions of the $W^+$ associated production process FIG.~\ref{fig:my_label6}. The $Z'$ case exhibits the same sharp peak at $\cos\theta\simeq-1$, while the ALP panel shows the same smoother backward enhancement. This confirms that, for both muons originate from the new-particle $W^+W^-$ cascade, the angular shape is governed by the new-particle decay and is insensitive to whether the accompanying gauge boson is $W^+$ or $Z$.

\begin{figure}[ht]
		\centering
		\captionsetup[subfigure]{justification=centering}
		\begin{subfigure}[b]{0.45\linewidth}
			\centering
			\includegraphics[width=\linewidth]{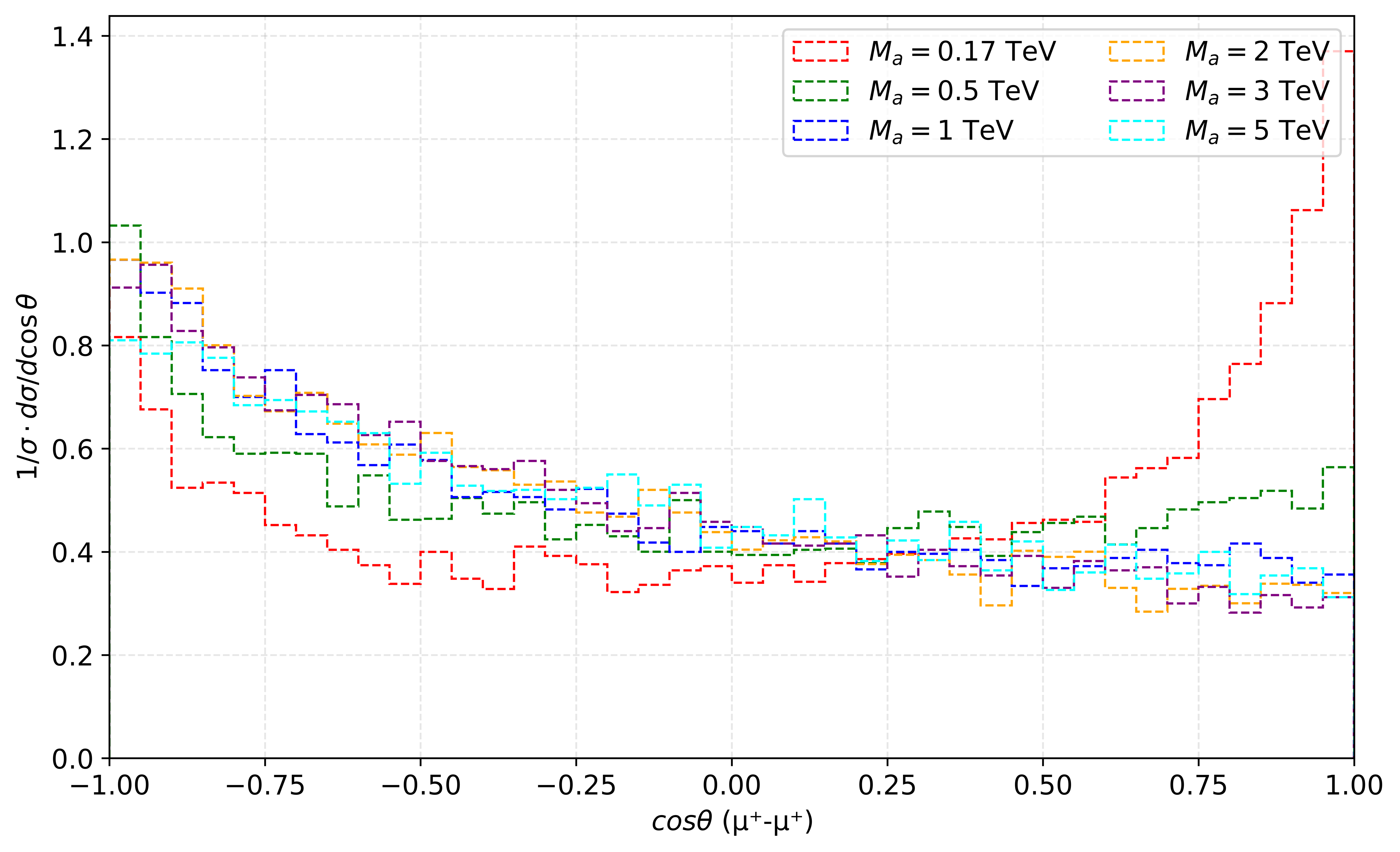}  
			\subcaption{}
			\label{fig:sub-a9}
		\end{subfigure}
		\hfill
		\begin{subfigure}[b]{0.45\linewidth}
			\centering
			\includegraphics[width=\linewidth]{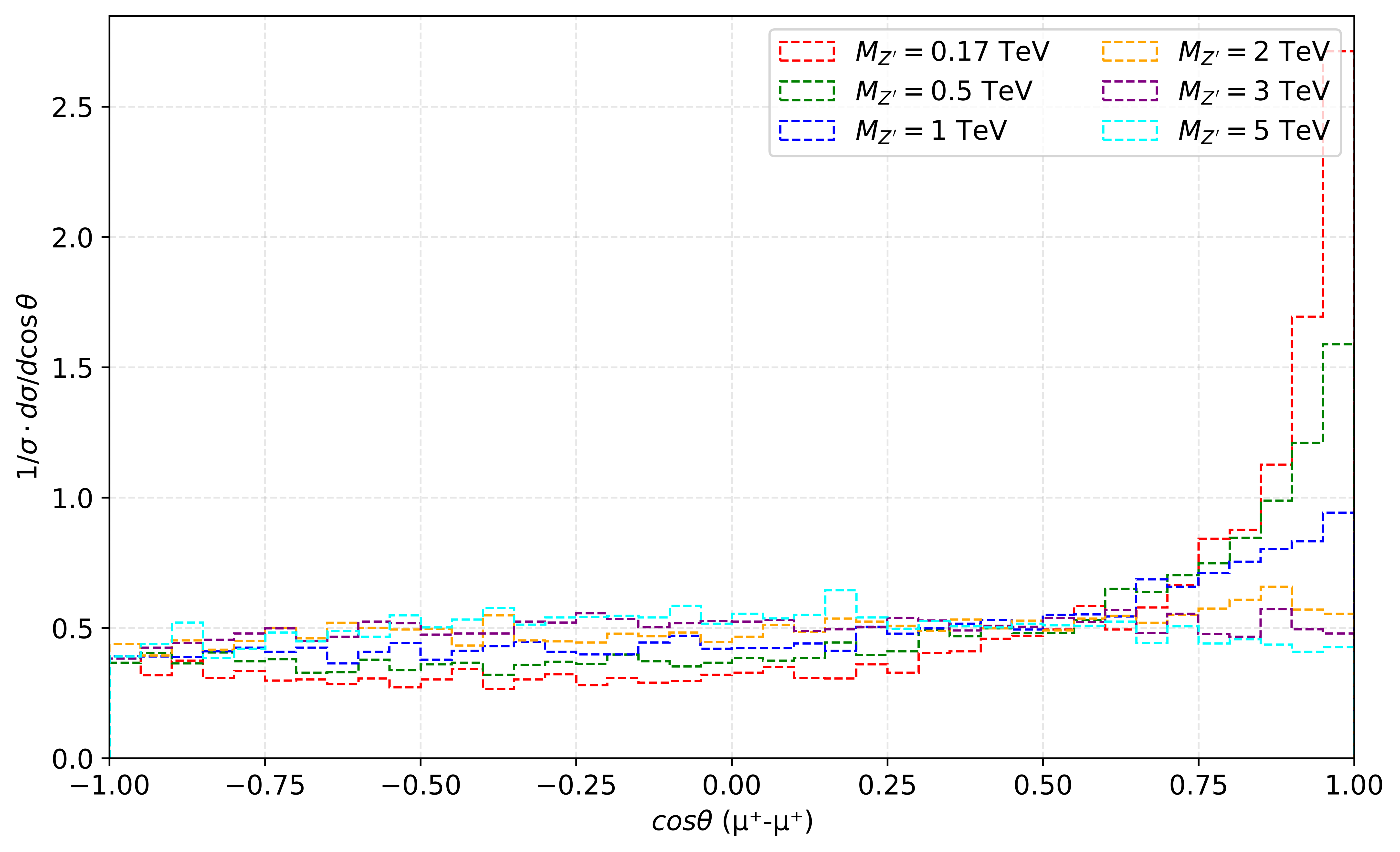}
			\subcaption{}
			\label{fig:sub-b9}
		\end{subfigure}
		\\[1ex]
		\begin{subfigure}[b]{0.45\linewidth}
			\centering
			\includegraphics[width=\linewidth]{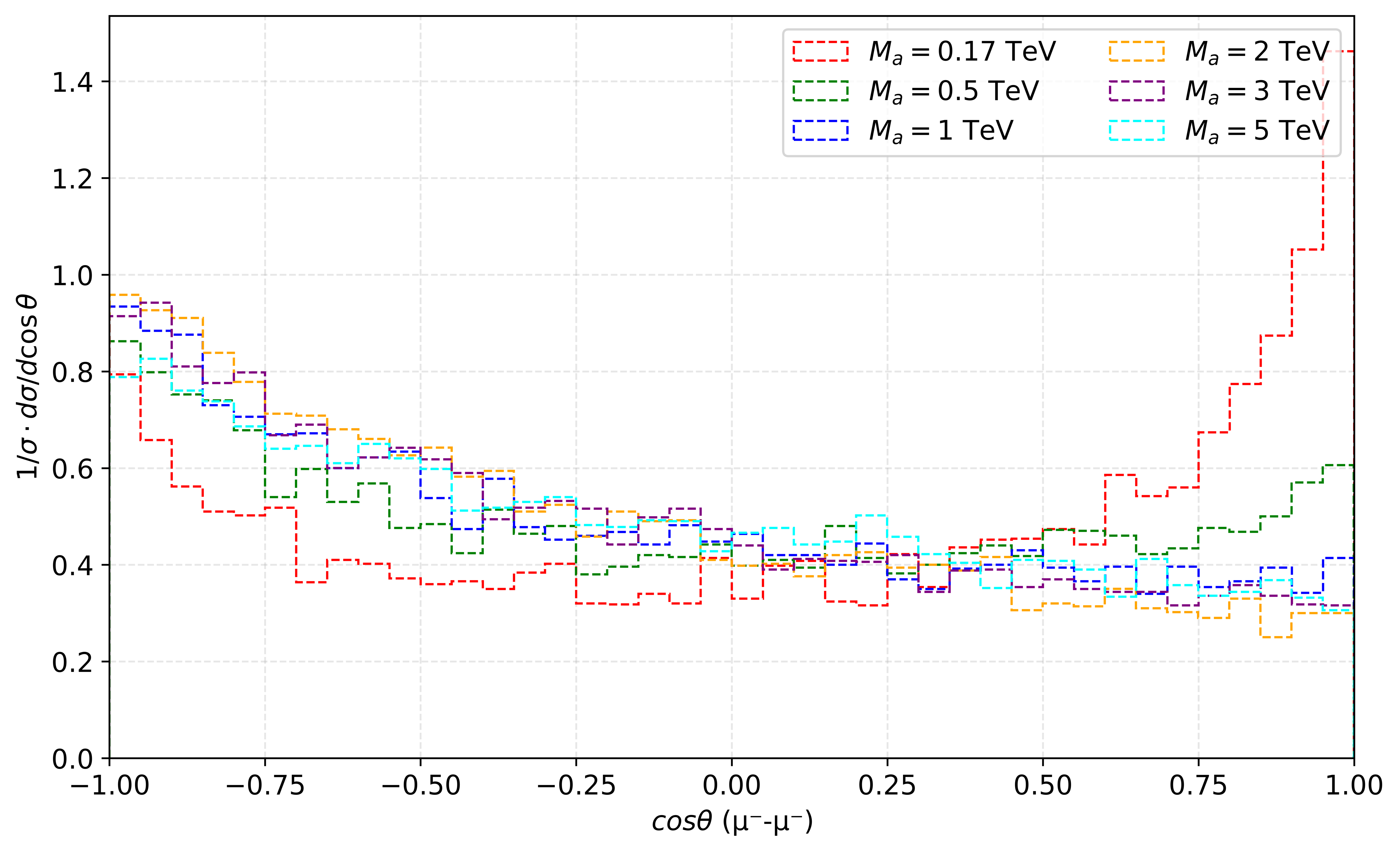}
			\subcaption{}
			\label{fig:sub-c9}
		\end{subfigure}
		\hfill
		\begin{subfigure}[b]{0.45\linewidth}
			\centering
			\includegraphics[width=\linewidth]{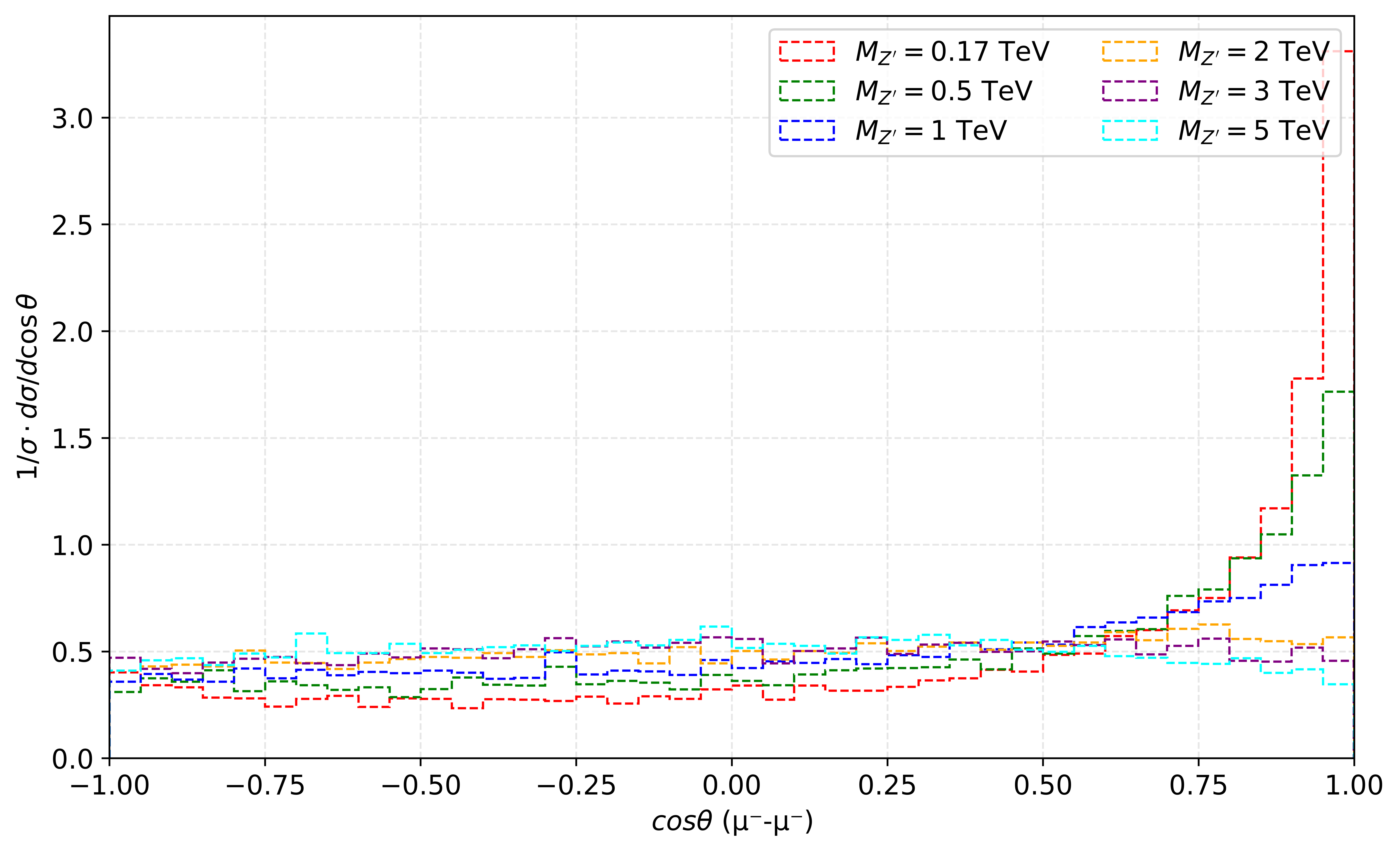}
			\subcaption{}
			\label{fig:sub-d9}
		\end{subfigure}
		\caption{Angular distributions of the final-state $\mu^+\mu^+$ pair from the process $pp \to Z Z'/a \to Z W^+ W^-\allowbreak\to \mu^+ \mu^+ \mu^- \mu^- \nu_\mu \bar{\nu}_\mu $ for different values of $M_a$ (a) and  $M_{Z'}$ (b) at $\sqrt{s}=14\ \text{TeV}$. The same case for  $\mu^-\mu^-$ pair is show in plot (c) and (d) for $M_a$ and $M_{Z'}$.}
		\label{fig:my_label9}
	\end{figure}

Having discussed the semileptonic channel, we now turn to the fully leptonic channel. Compared to the $W^+$ associated process, the $Z$ associated process can give rise to a four-lepton final state, which belongs to the fully leptonic channel and offers a cleaner signal. 
    FIG.~\ref{fig:my_label9} presents the normalized angular distributions of the same-sign muon pairs, $\mu^+\mu^+$ and $\mu^-\mu^-$, in the channel $pp \to Z Z'/a \to Z W^+ W^- \to \mu^+ \mu^+ \mu^- \mu^- \nu_\mu \bar{\nu}_\mu$.
	The left two panels (\subref{fig:sub-a9}) and (\subref{fig:sub-c9}) correspond to the ALP scenario, while the right two show the $Z'$ scenario. In this four-muon final state the same-sign muons are each composed of one muon from the $Z \to \mu^+\mu^-$ decay and one from the $W^+ \to \mu^+\nu_\mu (W^- \to \mu^-\bar{\nu}_\mu)$ decay, and we analyze the $\mu^+\mu^+$ and $\mu^-\mu^-$ distributions separately.
    We find that the normalized angular distributions for $\mu^+\mu^+$ and $\mu^-\mu^-$ are nearly identical, with the results for the two same-sign configurations essentially coinciding. This reflects the charge conjugation symmetry of the process at the parton level.
	Moreover, the same-sign muon angular behavior in this associated production channel closely resembles that of the $W^+$ associated production channel discussed above, indicating that the conclusions drawn for the process in FIG.~\ref{fig:my_label4} and ~\ref{fig:my_label5} remain valid for this process. These distributions allow us to efficiently distinguish vector particles from pseudoscalar particles in the new physics model.
    Similarly, the $A_{FB}$ values for the $Z$ associated production processes are summarized in TABLE~\ref{tab:U3}. When considering fully leptonic decays, most kinematic configurations can be used to efficiently discriminate between the $Z'$ and ALP particles.  
    
\begin{table}[H]
\centering
\setlength{\tabcolsep}{10pt}
\begin{tabular}{|c|c|c|c|c|c|c|c|}
\hline
 \makecell{$M_a$ \\ {[TeV]}}
  & \multicolumn{3}{c|}{$A_{FB}$}
  & \makecell{$M_{Z'}$ \\ {[TeV]}}
  & \multicolumn{3}{c|}{$A_{FB}$} \\[-2.0ex]
\cline{2-4} \cline{6-8}
  & D & E+ & E- & & D & E+ & E- \\
\hline
0.17 &  0.81 &  0.13 &  0.13 & 0.17 &  0.52 &  0.38 &  0.43 \\
0.5  &  0.32 & -0.099 & -0.096 & 0.5  &  0.055 &  0.27 &  0.30 \\
1    & -0.17 & -0.22 & -0.21 & 1    & -0.29 &  0.17 &  0.20 \\
2    & -0.66 & -0.28 & -0.30 & 2    & -0.72 &  0.068 &  0.076 \\
3    & -0.85 & -0.27 & -0.27 & 3    & -0.90 &  0.025 &  0.02 \\
5    & -0.81 & -0.22 & -0.20 & 5    & -1.00 &  0.003 & -0.015 \\
\hline
\end{tabular}
\renewcommand{\thetable}{\Roman{table}}
\caption{The $A_{FB}$ values of the muon pairs from the process $pp \to Z Z'/a \to Z W^+ W^-$ for different values of $M_a$ and $M_{Z'}$ at $\sqrt{s}=14\ \text{TeV}$. Column D corresponds to the process shown in FIG.~\ref{fig:my_label8}; columns E$+$ and E$-$ correspond to the $\mu^+\mu^+$ and $\mu^-\mu^-$ angular distributions in FIG.~\ref{fig:my_label9}, respectively.}
\label{tab:U3}
\end{table}
	\section{Summary}\label{sec:sum}
    Heavy vector and pseudoscalar particles are frequently predicted in the new physics theories. Studying the interactions between these two types of particles and gauge bosons constitutes an important part of understanding new physics models. In this work, we take two representative candidates as benchmarks: vector boson $Z'$ from a $U(1)_X$ gauge extension of the SM and the pseudoscalar axion-like particle $a$ associated with a Peccei--Quinn-like spontaneously broken global $U(1)$ symmetry. 
    We focus on a photophobic ALP to isolate its interactions with electroweak gauge bosons at high mass scales, with a large branching ratio of $a\to W^+W^-$
    . The associated production process shares the same final-state particles as the $Z' \to W^+W^-$ decay process.
    To investigate the properties of $Z'WW$ and $aWW$ interaction,  we systematically study the production of $Z'$ and $a$ at the $pp$ collider with collision energy of $14~\mathrm{TeV}$ and $100~\mathrm{TeV}$.
    
    The production of single resonance states at hadron colliders is an effective process to study the $Z'WW$ interaction.  The cross section for $pp \to Z'\to W^+W^-$ is $1.08\times 10^{-5}~\mathrm{pb}$ with $W$ boson semileptonic decaying at $\sqrt{s}=14~\mathrm{TeV}$ and $M_{Z'} = 5~\mathrm{TeV}$. 
    To further compare the properties of $Z'$ and $a$ particles, we investigate the associated production processes of $W^+Z'/a$ and $ZZ'/a$. The cross section of the final states corresponding to the semileptonic decays of these processes can reach  $1.75\times 10^{-5}~\mathrm{pb}$ and $1.00\times 10^{-5}~\mathrm{pb}$ for $\sqrt{s}=100~\mathrm{TeV}$ and $M_{Z'} = 5~\mathrm{TeV}$, and the cross section is $3.13\times 10^{-5}~\mathrm{pb}$ and $1.35\times 10^{-5}~\mathrm{pb}$ for $M_{a} = 5~\mathrm{TeV}$.

    The kinematic distributions of final state charged leptons constitute a powerful observable to study the interactions between the $Z'/a$ boson and gauge bosons. We define the angular distributions of the two final state leptons and present the muon angular distributions for different parent particles in the  $pp \to W^+ Z'/a \to W^+ W^+ W^-$ process. Our results show that the normalized differential cross section with respect to the final state angular distribution varies noticeably with the mass of the new particle. The angular distributions of same-sign and opposite-sign charged-lepton pairs, where the two leptons originate respectively from the associated produced $W$ boson and decayed from $W$ boson of the $Z'/a$ decay products, exhibit clear discrimination between the $Z'$ boson and the $a$ particle of identical mass. In particular, the forward–backward asymmetry $A_{FB}$ defined from the angular distributions has opposite signs for these two particles. Furthermore, for the fully leptonic decay channel in the associated $ZZ'/a$ production process, the final-state distributions of same-sign charged leptons also provide a distinct separation between the $Z'$ boson and the $a$ particle. For example, at $M_{Z'/a} = 3~\mathrm{TeV}$, in the process $pp \to Z Z'/a \to Z W^+ W^- \to \mu^+ \mu^+ \mu^- \mu^- \nu_\mu \bar{\nu}_\mu$, the $A_{FB}$ of the $\mu^+\mu^+$ pair is $0.025$ for the $Z'$ and $-0.27$ for the ALP.
    In summary, through systematic cross-section calculations and angular distribution analyses, this work not only provides the study for the $Z'$ boson and the ALP coupling with the gauge bosons, but also reveals the 
    distinguishing characteristics of the $Z'$ and the ALP to verify or exclude these two classes of new-physics models.
	\section*{Acknowledgment}

	This work is supported by Natural Science Foundation of Shandong Province under Grant No.ZR2026MS0091 and No.ZR2026QC1250, National Natural Science Foundation of China under Grant No.12605200 and No.12505112, and University of Jinan Disciplinary Cross-Convergence Construction Project 2024 (XKJC-202404).
    
	\section{References}
	\bibliographystyle{apsrev4-2}
	\bibliography{ref}

\end{document}